\documentclass[seceq]{ptptex}

\usepackage{graphicx,color} 
\usepackage{amssymb} 
\usepackage{latexsym}

\title{
Iterative Machine Learning  for  Output Tracking
}

\author{Santosh \textsc{Devasia}\footnote{S. Devasia is with the Mechanical Engineering Department, U. of Washington, Seattle,
WA 08195-2600 USA e-mail: devasia@uw.edu (see http://faculty.washington.edu/devasia/).}
}

\inst{
U. of Washington, Seattle, WA 98195-2600, USA
}

\abst{
This article develops   iterative machine learning (IML)  for output tracking.
The input-output data generated during iterations to develop the model used in the iterative update. 
The main contribution of this article to propose the use  of kernel-based 
machine learning to  iteratively update both the model and the model-inversion-based input simultaneously. 
Additionally,  augmented inputs with persistency of excitation are proposed  to promote 
learning of the model during the iteration process.  The proposed approach is illustrated with a simulation example.
}

\newcommand {\qed}{\hfill\mbox{\rule[0pt]{1.3ex}{1.3ex}}}
\newtheorem{rem}{\textbf{Remark}}
\newtheorem{lemma}{\textbf{Lemma}}

\newtheorem{assumption}{\textbf{Assumption}}

\newtheorem{condition}{\textbf{Condition}}

\newcommand{\jw}{(\omega)}

\DeclareMathOperator{\sgn}{sgn}

\begin{document}

\maketitle

\section{Introduction}
\label{section_introduction}
Iterative learning methods, initially developed in  e.g.,~\cite{arimoto_bettering_84, craig_84,AtkesonCG_1986},  
improve the  output-tracking performance by correcting the input based on the measured tracking error. 
For example, iterative  control  has led to some of the highest precision for output tracking, e.g., in scanning probe microscopy as demonstrated 
in, e.g.,~\cite{Bechhoefer_ACC_08,clayton_RSI_07,schitter_04,Zou_ACC_08,Leang_Mechatronics_06,Zou_CST_07,Chaffe_Pao}. 
Note that sets of learned trajectories can be used to enable tracking of other trajectories, e.g.,  by designing 
the feedforward control input using pre-specified basis functions~e.g., using polynomial functions~\cite{Meulen_asme_06} 
or rational functions of the reference trajectory~\cite{Oomen_15}. Similarly, new desired output can be generated 
by considering different combinations of previously-learned  output segments~\cite{Sandipan_tomi} and the feedforward 
can be represented using  radial basis functions that can be optimized for a range of task parameters~\cite{Gorinevsky_97}. 
This article proposes a kernel-based machine learning approach to use augmented inputs to 
iteratively learn not just the inverse input $u_{d}$ needed to track a specified output $y_{d}$ but to also use the data acquired during the iteration process to  
to estimate both (i)~the model $\hat{g}$
(and its inverse) for the control update, as well as (ii)~the model uncertainty needed to establish bounds on 
the iteration gain for ensuring  tracking-error reduction..

%

\vspace{0.05in}
Conditions for convergence of iterative methods have been well studied in literature, e.g.,~\cite{Moore_93, Tien_04_ACC, hatonen_04,szuchi_05,Moore07_review,jayati01_ASME,Bristow_08,jayati02,Peng_Tomi_16,Bing_freeman_16}. 
For example, the need to invert the system  $g$ to find the perfect input $u_d$ 
for tracking a desired output $y_d$ has 
motivated the use of the inverse $\hat{g}^{-1}$ of the known model $\hat{g}$ of the system $g$ in  early iterative control 
development~\cite{AtkesonCG_1986}.  
Since convergence depends on the size of modeling error,  improvements 
of the model through parameter adaptation with data acquired during the iteration was studied in, e.g., \cite{Norrlof_02} for robotics application using 
a discrete-time implementation. Here, for each iteration step,  the sampled input vector is mapped to the sampled output vector through a   
lower triangular matrix map. A stochastic version using such a lower-triangular map has been studied in~\cite{Butcher_08}. 
Even in the ideal case with no modeling uncertainty, the inverse of this matrix map leads to a stable inverse only if the system is minimum-phase (i.e., no zeros on the right hand side of the complex plane), e.g.,~\cite{Moore07_review,Sandipan_14}. This restriction to minimum-phase systems also applies to the use of 
input-output data to estimate models that enhance portability of the data-based learning to other output trajectories, e.g.,~\cite{Norrlof_02,Sandipan_14}.
The extension of iterative learning control   for nonminimum phase systems  using the noncausal inverse was initially proposed in~\cite{jayati01_ASME}. 
The  frequency domain implementation of the noncausal inverse was studied in, e.g.,~\cite{Tien_04_ACC,hatonen_04}, and discrete-time 
implementation was developed in~\cite{Oomen_14}, where  noncausality was allowed by 
using  full matrices for the input-output map  and  rational basis functions were used 
to enable portability between different trajectories in~\cite{Oomen_15_ACC}. 
%
Convergence  to the desired output can be  guaranteed (with the frequency domain approach)  if 
the phase uncertainty in the model is less than $90$ degrees and the iteration gain is sufficiently small~\cite{Tien_04_ACC,szuchi_05}. Such 
dependence on the phase uncertainty was also developed using a discrete Fourier transform approach in~\cite{hatonen_04,freeman_ASME}.

Convergence cannot be guaranteed in  regions of the frequency domain where   the phase uncertainty in the model $\hat{g}$ is greater than $90$ degrees. 
Frequency  regions where convergence cannot be guaranteed can be reduced by using the input-output data generated during the iteration procedure~\cite{zou_iterative_2013}.   
In particular, more recent model-less approaches use the input-output data from previous iteration steps to avoid the need to model the system explicitly and improve the convergence of the iterative approach~\cite{zou_iterative_2013,Wang_zou_iterative_2015}. Nevertheless, such input-output data might have substantial error at some frequencies where the signal to noise ratio is small. 
Kernel-based Gaussian process regression (GPR)~\cite{Rasmussen_06} is well suited to such function-estimation from noisy data. 
This motivates the main contribution of the article   \textemdash \  the use of a kernel-based iterative machine learning (IML)  
approach to predict the magnitude and phase response of the 
system as functions of frequency from the measured input-output data. 
An added advantage  is that the IML approach also yields the anticipated model uncertainty, which can be used to design the iteration law using previous results from  \cite{Tien_04_ACC,hatonen_04,szuchi_05}. 


\vspace{0.1in}
A second contribution of this article is to propose 
the use of additional input to the iteration law  for persistency of excitation 
to  enhance the model learning. 
It is shown in the article that convergence cannot be guaranteed 
if the size of the model is small compared to the size of the error in the model data, e.g.,  due to noise in the output.  
The effect of the noise in output measurement can be reduced by  increasing the size of the output by using larger input. 
This increase in the output-to-noise ratio, i.e., the ratio of 
measured output to known input in the frequency domain,   reduces the error in the  model data  and improves the 
performance of the iteration procedure. 
Therefore, this article proposes the injection of additional input into the iteration law whenever the the input size falls below a 
threshold value.  This persistence of excitation leads to  smaller noise in the measured system-model data with the proposed approach. 
(The effect of this input augmentation is removed from the measured output, using the estimated model, before the input is 
updated at the next iteration step.) Note that similar use of input augmentation to ensure persistency of excitation  is commonly used in adaptive control, e.g.,~\cite{astrom_param}. 
The improved model  can be used to better infer the inverse input for new 
output trajectories. In this sense, the proposed approach enhances the portability of the frequency-domain iterative learning control. 

\vspace{0.1in}
The proposed use of additional input to improve model learning during the iteration process, is general in the sense that it can be used 
with other modeling methods. For example, in the current work a non-parametric Gaussian process regression (GPR)  is proposed to predict the system response   from the measured input-output data~\cite{Rasmussen_06}. However, the proposed  model update during the iteration steps can also be applied using parametric methods for modeling, e.g., see  recent review in~\cite{Pillonetto2014657}.  The models could then be inverted (potentially, in the time domain) during iterations for input update --- in contrast, the proposed approach directly measures and stores the frequency data of the model (and hence its inverse). The GPR provides the  smoothing in the presence of noisy data.  
Similarly, there is flexibility in the selection of the kernel used for estimating the models. While the more common smooth squared exponential (SE) kernel is used in this article, the Matern class of kernels could be used for systems with sharper features in the frequency response, e.g., for underdamped systems~\cite{Rasmussen_06} and the approach could be applied to 
complex-valued kernels in the frequency domain~\cite{Tortosa_Fuentes}. 
The concept of model update in the frequency domain proposed here could  be used with spatial-domain iterations, e.g.~\cite{Landers_Bristow_13,Kira_16}, with model identification methods
using repetitive trajectories~\cite{alleyne_16}, and  with the stable spline kernel for machine learning in the time domain~\cite{Pillonetto2014657} that guarantees bounded-input-bounded-output stability of the resulting models. Finally, the proposed approach can be used to speed up the learning of the different segments in segmented iterative control approaches, e.g.,~\cite{Sandipan_tomi}.

\vspace{0.1in}
The paper begins with problem formulation in Section~\ref{section_problem_formulation}, where the standard model-inversion-based iterative control approach is briefly reviewed followed by clarification of the research problem and the proposed solution approach using kernel-based machine learning in Section~\ref{Kernel_section}. Convergence conditions are then developed in Section~\ref{section_convergence_conditions_IMA}, which are used to redesign the iteration law to promote persistency of excitation. 
Additionally, the overall IML algorithm is presented at the end of Section~\ref{section_convergence_conditions_IMA}.  Simulation results and discussion illustrating the proposed approach are in Section~\ref{sec_results} followed by the conclusions in Section~\ref{sec_conclusions}.

\vspace{0.1in}
\section{Problem formulation}
\label{section_problem_formulation}
This section begins by discussing the system and limits of the model-inversion approach due to 
modeling error, followed by background on frequency-domain iterative control to correct for modeling error. 
Convergence conditions and approaches to reduce the modeling error from data are presented, followed 
by the research-problem statement. 

\subsection{Model-based inverse feedforward}
Given a desired output $ y_d$, model inversion can be  used to find the  feedforward input $u = u_d$ 
that achieves the desired output $y = y_d$ for a linear system of the form 
\begin{equation}
\label{system_eq}
y(s) =  g(s) u (s)
\end{equation}
as 
\begin{equation}
\label{inverse_system_eq}
\begin{aligned}
u_{d}\jw  ~ & =   g^{-1}\jw y_{d}\jw  
~= [ {a} \jw  + j  {b}\jw ]^{-1} y_{d}\jw
\end{aligned}
\end{equation}
with the 
value of the model $\hat{g}$ evaluated on the imaginary axis of the complex plane defined as 
$\hat{g} \jw  =   \hat{g} (s) {\Huge{|}}_{s = j\omega}$   and  $j = \sqrt{-1}$.

\vspace{0.1in}
\begin{assumption}[System properties]
\label{assumption_system_and_model_properties}
The system  $g$ is not identically zero (i.e., it is non-trivial), is stable, 
and has hyperbolic zero dynamics, i.e., all zeros have a nonzero real
parts. 
\end{assumption}

\vspace{0.1in}
Note that if the known system model  $\hat{g}\jw$ has error at frequency $\omega$, i.e., $\hat{g}\jw \ne g\jw$, then the inverse 
input $\hat{u}_d\jw$ 
\begin{equation}
\label{inverse_system_model_eq}
\hat{u}_d\jw =   \hat{g}^{-1}\jw y_{d}\jw 
\end{equation}
found using the model $\hat{g}\jw$  does not lead to 
exact output tracking of the desired output $y_d\jw$, i.e., 
\begin{equation}
\label{tracking_inverse_system_model_eq}
 g\jw \hat{u}_d\jw  ~ =  g\jw \hat{g}^{-1}\jw y_{d}\jw~ \ne y_d\jw .
\end{equation}

\subsection{Fixed-model-based iterative control}
The tracking error caused by the modeling error, $\hat{g}\jw \ne g\jw$ at frequency $\omega$, can be corrected iteratively, e.g.,  as~\cite{szuchi_05,hatonen_04} 
\begin{equation}
\label{iteration_law_eq}
u_{k}\jw ~=  u_{k-1}\jw  + \rho_{k}\jw  \hat{g}_k^{-1} \jw \left[   y_d\jw - y_{k-1}\jw \right]
\end{equation}
provided $\hat{g}_k^{-1} \jw \ne 0 $ at frequency $\omega$, 
where at each integer iteration-step $k > 1$,  the input $u_{k}$ is computed using Eq.~\eqref{iteration_law_eq}
with  a real-valued scalar, frequency-dependent, iteration gain $\rho_{k}$ and the available system model $\hat{g}_k$ and applied to the 
system to find the output $y_{k}$. 
Note that  the term in the square bracket in Eq.~\eqref{iteration_law_eq}  represents the output error. 
The  iterative control in Eq.~\eqref{iteration_law_eq}  converges at frequency  $\omega$  if the 
modeling error is sufficiently small, as shown in~\cite{szuchi_05,hatonen_04}, and stated formally below. 

\vspace{0.1in}
\begin{lemma}[Output-tracking convergence]
\label{lemma_convergence_single_input}
With a finite initial input $u_{1}\jw$, 
a fixed iteration gain $ \rho_{k}\jw = \rho\jw $,  and non-zero system model $\hat{g}_k\jw = \hat{g}\jw \ne 0$  at frequency $\omega$ , 
 the  iterations  in Eq.~\eqref{iteration_law_eq} converges to the inverse input $u_{d}\jw$, i.e., 
\begin{equation}
\label{input_convergence_eq}
 \lim_{k \rightarrow \infty}  u_{k}\jw = u_{d}\jw ~=  g^{-1} \jw \left[   y_d\jw  \right] , 
  \end{equation}
which results in 
exact  tracking of the desired output $y_d\jw$, i.e., 
\begin{equation}
\label{exact_tracking_eq}
\lim_{k \rightarrow \infty} y_{k}\jw  =   \lim_{k \rightarrow \infty}   g\jw u_{k}\jw  =   y_d\jw ,  
 \end{equation}
if and only if the magnitude of the phase uncertainty  $\Delta_{p}$ in the model  $\hat{g}$ and   the iteration gain $\rho$ are 
sufficiently small  
\begin{equation}
\label{eq_conditions_single_input}
\begin{aligned}
| \Delta_{p} \jw | & <  \pi/2
\\
0 <   \rho\jw  &  <   
\frac{2\cos{\left[ \Delta_{p}\jw  \right]}}{\Delta_{m}\jw} , 
\end{aligned}
\end{equation}
where the  magnitude uncertainty $\Delta_{m}$ and the phase uncertainty $\Delta_{p}$ are defined by 
\begin{equation} 
\label{model_uncetainty_definition}
\frac{ \hat{g}^{-1}\jw } { g^{-1}\jw }  ~=  
\frac{ g\jw }{ \hat{g}\jw }   ~=  \Delta_{m} \jw  e^{j\Delta_{p}\jw}.
 \end{equation}
\end{lemma}

\vspace{0.01in}
\noindent
{\bf{Proof}:}
This follows from  Lemma~1 in~\cite{Tien_04_ACC,szuchi_05}. The phase condition is the same as in ~\cite{hatonen_04}.
Briefly, multiplying Eq.~\eqref{iteration_law_eq} by the system $g$ and subtracting from the desired output $y_d$ yields 
\begin{equation} 
\label{convergence_proof}
{ y_d\jw - y_k\jw }   ~=   \left(  1  -    \rho\jw \frac{ g\jw }{ \hat{g}\jw }  \right)     \left[  { y_d\jw - y_{k-1}\jw } \right], 
 \end{equation}
 which will tend to zero with increasing  iteration steps $k$  if 
 \begin{equation} 
\label{convergence_proof_2}
  \left|  1  -    \rho\jw \frac{ g\jw }{ \hat{g}\jw }  \right| =      \left|  1  -    \rho\jw \Delta_{m} \jw  e^{j\Delta_{p}\jw} \right|   < ~1.
   \end{equation}
   The lemma follows by squaring the right hand side of the above equation to obtain 
  \begin{equation} 
\label{convergence_proof_3}
\begin{aligned}
 1 - 2 \rho\jw  \Delta_{m}\jw \cos(\Delta_{p}\jw ) + \rho^2\jw \Delta_{m}^2\jw  < 1
 \end{aligned}
   \end{equation}
   and removing one from both sides to yield (since $ \Delta_{m}  > 0 $ because $ g\jw  \ne 0$, $ \hat{g}\jw  \ne 0$ ), 
  \begin{equation} 
\label{convergence_proof_4}
   \begin{aligned}
 \rho\jw  \left[ \rho\jw \Delta_{m}\jw  - 2 \rho\jw  \cos(\Delta_{p}\jw ) \right]  < 0 . 
 \end{aligned}
   \end{equation}
   Note that if the phase uncertainty is small, i.e., $|\Delta_{p}\jw|< \pi/2$, then $\cos(\Delta_{p}\jw ) > 0$ and  $\rho\jw$ needs to be positive  to satisfy Eq. \eqref{convergence_proof_4}. 
 However, if the phase uncertainty $|\Delta_{p}\jw| $ is greater than  $ \pi/2$,  then a fixed  $\rho\jw$ cannot satisfy Eq. \eqref{convergence_proof_4} for all uncertainties since 
   $\cos(\Delta_{p}\jw)$  can potentially be positive or negative. Finally, when  $|\Delta_{p}\jw|  = \pi/2$,  the left hand side is positive. 
\qed

\vspace{0.1in}
\begin{rem}[Small phase error]
\label{rem_phase_error}
The above condition implies that tracking errors reduce at each iteration step, and the input iterations will converge to the 
desired inverse input as in Eq.~\eqref{input_convergence_eq} at each frequency $\omega$ if  and only if the iteration gain $\rho\jw$  is chosen to be sufficiently small, provided 
the phase error $\Delta_{p}\jw$ is less than $90$ degrees~\cite{Tien_04_ACC,szuchi_05,hatonen_04}. 
\end{rem}
x


\subsection{Data-based model update for iterative control}
The inverse model $\hat{g}_k^{-1}\jw$ in the iterative control in Eq.~\eqref{iteration_law_eq}, used to compute the input $u_{k}\jw$ at iteration step $k$, 
can be estimated 
from the previously computed input $u_{k-1}\jw$ and the corresponding output $y_{k-1}\jw$  as 
\begin{equation}
\label{model_update_qingze_eq}
\hat{g}_{k}^{-1}\jw ~= 
 \frac{u_{k-1}\jw}{y_{k-1}\jw} 
 \end{equation}
if $ y_{k-1}\jw \ne 0$.
Provided  the noise in the estimation of the model $\hat{g}_{k}\jw$ ($k > 1$) is small, the 
above approach can be used with an iteration gain of $\rho_k\jw =1$ to modify the 
iterative input-update law in Eq.~\eqref{iteration_law_eq} to~\cite{zou_iterative_2013} 
\begin{equation}
\label{iteration_law_qingze_eq}
u_{k}\jw ~ = 
\left\{ 
\begin{aligned}
 &  \frac{u_{k-1}\jw}{y_{k-1}\jw}  y_d\jw   & \quad  {\mbox{if}}~ y_{k-1}\jw \ne 0 \\ 
& 0  & \quad  {\mbox{otherwise}} 
\end{aligned}
 \right. 
\end{equation}
and the initial input at $k=1$ is considered to be 
\begin{equation}
\label{initial_input_eq}
u_1\jw = \alpha y_d\jw
\end{equation}
 where $\alpha$ is a constant 
that can be chosen, e.g., to be the inverse of the estimated DC gain of the system. The tracking error 
can be made small if the noise in the measurements is small, as shown in~\cite{zou_iterative_2013}.   

\vspace{0.1in}
\begin{rem}[Model filtering]
\label{rem_model_filtering}
The estimated inverse model 
$\hat{g}_k^{-1}\jw$ in the iterative control in Eq.~\eqref{iteration_law_eq} at iteration step $k$  can be replaced by  a weighted average 
from  all available estimates from iteration steps $k_i \le k$  to reduce the effect of noise~\cite{Wang_zou_iterative_2015}. 
\end{rem}

\subsection{The research problem}
There are two main issues with current  frequency-domain iterative approaches, e.g.,~\cite{szuchi_05,zou_iterative_2013,hatonen_04}. 
The first issue is that while these frequency-domain methods iteratively find the inverse input $u_d$ to achieve a desired output $y_d$, they do not directly improve the ability to track a new output trajectory $y_{d,2}$, with potentially different frequency content. 
Secondly, current approaches do not  improve estimates of the uncertainties ($\Delta_m, \Delta_p$) 
in the  model $\hat{g}$. Note that   lower model uncertainties  
can allow the use of larger iteration gains according to Eq.~\eqref{eq_conditions_single_input} and can lead to faster convergence. 

\vspace{0.1in}
The research problem is to iteratively learn the system model $\hat{g}$  (and therefore, its inverse $\hat{g}^{-1}$) and estimate the model uncertainties ($\Delta_m, \Delta_p$) 
while iteratively learning the inverse input $u_d$ needed to track a desired 
output $y_d$. 

\vspace{0.1in}
\section{Proposed machine learning of  model}
\label{Kernel_section}
A kernel-based machine learning approach is proposed in this section to learn the system model  $\hat{g}$ during the iteration process. 
The Gaussian process assumption is clarified first in Subsection~\ref{Subsection_GP_assumption}, followed by background on the Gaussian process regression (GPR) approach to estimate a general function $f$ in Subsection~\ref{Subsection_GPR}, and lastly, by  the application 
of the GPR  to estimate the system model in Subsection~\ref{Subsection_GPR_application}.  

\subsection{Model is assumed to be a Gaussian process}
\label{Subsection_GP_assumption}
In the following,  the real $\hat{a}$ and imaginary $\hat{b}$ components  of the model $\hat{g}$ are considered to be 
independent  real-valued Gaussian processes as in~\cite{Perez_split_Gaussian}.

\vspace{0.1in}
\begin{assumption}[Gaussian process]
\label{assumption_GP_properties}
The  real $\hat{a}$ and imaginary $\hat{b}$ components of the  system model $\hat{g}$
are considered to be  zero-mean, real-valued,  independent Gaussian processes with covariance functions $k_a(\omega,\omega^\prime)$ 
and $k_b(\omega,\omega^\prime)$ respectively, i.e., 
\begin{equation}
\label{inverse_GPs_eq}
\begin{aligned}
\hat{a}\jw & =  {\mathcal{R}} [ g\jw ]   \sim  GP\left(0, k_a(\omega,\omega^\prime)\right)  \\ 
\hat{b}\jw & =    {\mathcal{I}} [  g\jw ]  \sim  GP\left(0, k_b(\omega,\omega^\prime)\right) , 
\end{aligned}
\end{equation}
where  the measured real $m_a$ and imaginary $m_b$  components  are  given by 
\begin{equation}
\label{inverse_GPs_eq_2}
\begin{aligned}
m_a& =  \hat{a}\jw + \epsilon_a    \\ 
m_b & =  \hat{b}\jw + \epsilon_b    \\ 
\end{aligned}
\end{equation}
with additive, zero-mean, independent identically distributed Gaussian noise $ \epsilon_a,  \epsilon_b $ with 
variance $\sigma_{a}^2, \sigma_{b}^2$. 
\end{assumption}

\vspace{0.1in}
\begin{rem}[Lack of knowledge about the model]
\label{rem_model_filtering}
The Gaussian process  has  zero mean in Assumption~\ref{assumption_GP_properties} ---  
the estimated model $\hat{g}\jw$ tends to zero at frequencies $\omega$  far from the frequency region where 
data is available.  
The size of the acceptable modeling uncertainty (for using inversion-based input update) 
depends on the estimated model size, as discussed  in the next section. 
\end{rem}

\vspace{0.1in}
 \begin{rem}[Related real and imaginary components]
 \label{rem:relation_real_imag}
  The real and imaginary components of a causal linear transfer function ${g}$ are related to each other by the Kramers-Kronig relations, e.g.,~\cite{Bechhoefer_11_AJP}.
However, this relation between the real and imaginary components is not pointwise in frequency. For example, the real part $\hat{a}\jw$ at a frequency $\omega$ depends on the  complex part $\hat{b}(\cdot)$ over the 
entire frequency domain. 
The approach used here to separately estimate the real  $\hat{a}$  and imaginary  $\hat{b}$  components  of the model $\hat{g}$  is conservative and over-predicts the uncertainty. Therefore, it can lead to a smaller  iteration gain $\rho$  with slower convergence. 
\end{rem}

\subsection{Background: Gaussian process regression}
\label{Subsection_GPR}

\vspace{0.1in}
The data-based estimation of the expected value general function $f(\cdot)$  with $n$ measurements $m \in {\cal{R}}_{n} $ at 
frequencies $\Omega \in {\cal{R}}_{n}$ at frequency $\omega \in {\cal{R}}$  using Gaussian process regression (GPR) is stated formally 
in the following  lemma, e.g.,~\cite{Rasmussen_06}. 

\vspace{0.1in}
\begin{lemma}[Machine learning]
\label{lemma_machine_learning}
Let $f$ be a zero-mean real-valued Gaussian process over the frequency space 
\begin{equation} 
f(\omega)  \sim  GP\left(0, k(\omega,\omega^\prime)\right)
\end{equation}
with 
covariance function $k(\omega,\omega^\prime)$, and measurements  
 $m \in {\cal{R}}_{n} $ given by 
\begin{equation}
\label{general_GPs_eq}
m ~ =  f\jw + \epsilon    
\end{equation}
with additive, zero-mean, independent identically distributed Gaussian noise with variance $ \sigma^2$. 
Then, the 
prediction ${f_\star}$  at  any prediction  frequency $\omega \in {\cal{R}}$, given $n$ measured   data 
\begin{equation} 
{\bf{m}} =
\left[ m_1  ~ m_2  ~ \hdots   ~ m_{n}  \right]^T
\end{equation} 
(with the superscript $T$ indicating matrix transpose) 
at  frequencies  
\begin{equation} 
\Omega = 
\left[ \omega_1  ~ \omega_2  ~ \hdots   ~ \omega_{n}  \right]^T, 
\end{equation}
is Gaussian~\cite{Rasmussen_06} 
\begin{equation}
\label{general_GP_regression_eq}
\begin{aligned}
{f_{\star}} {\Huge{|}} \Omega, {\bf{m}}, \omega ~  \sim  {\mathcal{N}}\left(\bar{f_\star },  \sigma_{f_\star}^2   \right), 
\end{aligned}
\end{equation}
where the predicted mean $\bar{f_\star }$ and  variance $\sigma_{f_\star}^2 $ at any prediction frequency $\omega$  are given by 
\begin{equation}
\label{general_GP_regression_mean_cov_eq}
\begin{aligned}
\bar{f_\star }\jw ~ & ~= K(\omega, \Omega) \left[ K(\Omega, \Omega)  + \sigma^2 I   \right]^{-1}  {\bf{m}}  \\
\sigma_{f_\star}^2\jw ~ & ~= K(\omega, \omega)   \\
~ & \quad  -K(\omega, \Omega)\left[ K(\Omega, \Omega)  + \sigma^2 I   \right]^{-1} K(\Omega, \omega)      
\end{aligned}
\end{equation}
with $K(\Omega, \omega)$ denoting the   covariances evaluated at all pairs of measured 
frequencies  $\Omega \in {\cal{R}}_n $ and the prediction frequency $\omega  \in {\cal{R}} $. 
The other covariance matrices $K (\cdot, \cdot)$ are defined similarly. 

\end{lemma}

\vspace{0.05in}
\noindent
{\bf{Proof}:}
See, e.g., Chapter~2 in~\cite{Rasmussen_06}.
\qed

\subsection{Estimation of model from data}
\label{Subsection_GPR_application}
The data-based estimation of the real $\hat{a}$ and imaginary $\hat{b}$ components of the  system model $\hat{g}$
using GPR from the above Lemma~\ref{lemma_machine_learning}, e.g.,~\cite{Rasmussen_06} is described below.

\vspace{0.1in}
At iteration step $k$, given all the measured real  ${\bf{m}}_{a,k}$ and imaginary ${\bf{m}}_{b,k}$ components at frequencies $\Omega_k$,  
the predictive means $ \bar{a}_{\star,k}\jw,  \bar{b}_{\star,k}\jw  $  and variances $\sigma_{a_{\star,k}}^2\jw, \sigma_{b_{\star,k}}^2\jw$ 
of the real  $a\jw$ and imaginary   $b\jw$ components at prediction frequency $\omega$ can be obtained 
by setting
\begin{equation}
\label{output_fit_eq}
{f_\star }  = {z}_{\star,k},  \quad \sigma_{f_\star }  = \sigma_{{z}_{\star,k}}, \quad {\bf{m} }  = {\bf{m}}_{z,k} , \quad      \Omega = \Omega_{k} 
\end{equation}
in Eq.~\eqref{general_GP_regression_eq} and Eq.~\eqref{general_GP_regression_mean_cov_eq} of the above Lemma~\ref{lemma_machine_learning} where the subscript $z$ is replaced by either $a$ or $b$  depending on whether the real or imaginary component  is being predicted. 

\vspace{0.1in}
 The resulting  model $\hat{g} _k$ at the $k^{th}$ iteration step  is given by 
\begin{equation}
\label{model_update_ML_eq}
\begin{aligned}
\hat{g}_{k}\jw  & ~= [ \bar{a}_{\star,k} \jw  + j  \bar{b}_{\star,k}\jw ]
 ~:= [ \hat{a}_{k} \jw  + j  \hat{b}_{k}\jw ] .
\end{aligned}
 \end{equation}

 
\vspace{0.1in}
\section{Convergence with machine learned model}
\label{section_convergence_conditions_IMA}
Convergence of the iterative approach with the estimated model  from machine learning 
can be quantified in terms of the size of the model uncertainties in its real $\hat{a}_k$ and imaginary $\hat{b}_k$ components, as shown in this section.
This is  followed by a discussion on the augmentation of the iterative input law (at frequencies where the signal to noise ration is low) 
to reduce the impact of measurement noise when  estimating the model. This section concludes with the proposed algorithm.

\subsection{Convergence under bounded uncertainties}
\label{section_convergence_conditions}

\vspace{0.1in}
\subsubsection{Convergence conditions}
In the following lemmas, it is assumed that  the size of the  model uncertainty is  bounded.

\vspace{0.1in}
\begin{condition}[Bounds on model uncertainty]
\label{condition_uncertainty_bounds}
The deviations of the 
real and imaginary components $a, b$ of the system in Eq.~\eqref{inverse_system_eq} 
from the corresponding components $\hat{a}_k, \hat{b}_k$  of the model  in Eq.~\eqref{model_update_ML_eq} at iteration step $k$, 
are bounded as 
\begin{equation}
\label{model_uncertainty_ri_CI}
\begin{aligned}
|  a\jw - \hat{a}_k\jw |     &~~    \le  \Delta_{a,k} \jw   \\
 |  b\jw - \hat{b}_k\jw |     & ~~   \le  \Delta_{b,k} \jw  .  \\
\end{aligned}
 \end{equation}
\end{condition}

\vspace{0.1in}
\begin{rem}[Confidence intervals]
Predictive confidence intervals can be used as  bounds on the  model uncertainties  in Eq.~\eqref{model_uncertainty_ri_CI}. 
In this probabilistic setting, the chance of not converging can be made small, but the method does not guarantee that the tracking error will 
decrease. Nevertheless, if the model uncertainty falls outside of the specified confidence intervals at frequency $\omega$ 
and  the input $u_k\jw$  starts to grow, then   
the resulting output $y\jw$ will also  become 
large due to the hyperbolic-internal dynamics $g\jw \ne 0$ from Assumption~\ref{assumption_system_and_model_properties}. 
If the output error becomes large compared to the  noise 
in the measurement at some frequency $\omega$, then 
this  yields additional data  at frequency $\omega$ for improved model estimation, and input correction. 
\end{rem}

\vspace{0.1in}
\noindent
Conditions on the magnitude and phase uncertainties can be developed based on the uncertainties in the real and imaginary components of  the model, as shown below.

\vspace{0.1in}
\begin{lemma}[Bounds on phase and magnitude]
\label{lemma_uncertainty_bounds}
Under Condition~\ref{condition_uncertainty_bounds}, 
 the magnitude uncertainty $\Delta_{m,k}\jw$ and the phase uncertainty $\Delta_{p,k}\jw$,  similar to Eq.~\eqref{model_uncetainty_definition}, 
 \begin{equation} 
\label{model_uncetainty_definition_GPR}
 \Delta_{m,k} \jw  e^{j\Delta_{p,k}\jw} ~ = ~ \frac{g\jw}{\hat{g}_k\jw }  =  ~ \frac{a\jw + j b\jw} {\hat{a}_k\jw + j \hat{b}_k\jw }
 \end{equation}
satisfy 
\begin{equation}
|\Delta_{m,k}\jw | ~  \le  
\frac{|  \Delta_k\jw + \hat{g}_{abs,k}\jw  | }{  | \hat{g}_k\jw| }
\label{eq_M_k_condition_uncertainty}
\end{equation}
and
\begin{equation}
\cos[{\Delta_{p,k}}\jw] ~  \ge
\frac{ |\hat{g}_k\jw|^2  - \hat{g}_{abs,k}\jw.\Delta_k\jw}{
|\hat{g}_k\jw| ~~|\hat{g}_{abs,k}\jw  + \Delta_k\jw|
},
\label{eq_P_k_condition_uncertainty}
\end{equation}
where 
\begin{equation}
\begin{aligned}
 \Delta_k\jw  & =  \Delta_{a,k}\jw+ j\Delta_{b,k}\jw \\
\hat{g}_{abs,k}\jw  & = |\hat{a}_k\jw|  +  j |\hat{b}_k\jw| .
 \end{aligned}
\label{eq_P_k_condition_uncertainty_2}      
\end{equation}

 \end{lemma}

\vspace{0.01in}
\noindent
{\bf{Proof}:}
An expression for the magnitude $\Delta_{m,k}$ can be found from the definition  in Eq.~\eqref{model_uncetainty_definition_GPR}  
as 
\begin{equation}
\Delta_{m,k}\jw ~  =  
~ \frac{|g\jw|}{|\hat{g}_k\jw| } 
\label{eq_M_k_condition_uncertainty_derivation_1}
\end{equation}
and the condition in Eq.~\eqref{eq_M_k_condition_uncertainty}  follows since the 
numerator is maximized 
by selecting the maximial $|a\jw|, |b\jw|$  that satisfy the lemma's condition in Eq.~\eqref{model_uncertainty_ri_CI}, 
as in Eq.~\eqref{eq_M_k_condition_uncertainty}. 
The cosine of the phase angle between the model $\hat{g}_k\jw$ and the system $g\jw$ can be found using the dot product as 
\begin{equation}
\cos[{\Delta_{p,k}}\jw] ~  = 
\frac{ \hat{g}_k\jw.g\jw}{
|\hat{g}_k\jw| ~|g\jw|
} .
\label{eq_P_k_condition_uncertainty_2}
\end{equation}
The numerator in Eq.~\eqref{eq_P_k_condition_uncertainty_2} 
\begin{equation}
\begin{aligned}
\hat{g}_k\jw.g\jw   & =  \hat{a}_k\jw a\jw +  \hat{b}_k\jw b\jw   \\
&  =  \hat{a}_k^2\jw +  \hat{b}_k^2\jw   +\delta_{a,k}\jw\hat{a}_k\jw  +\delta_{b,k}\jw\hat{b}_k\jw, 
 \end{aligned}
\label{eq_P_k_condition_uncertainty_3}
\end{equation}
where 
\begin{equation}
\begin{aligned}
a\jw = \hat{a}_k\jw +  \delta_{a,k}\jw,  & \quad  &  b\jw = \hat{b}_k\jw +  \delta_{b,k}\jw  \\
| \delta_{a,k}\jw |   \le \Delta_{a,k}\jw, & \quad  & | \delta_{b,k}\jw|  \le \Delta_{b,k}\jw 
\label{eq_a_delta_def}
 \end{aligned}
\end{equation}
is minimized (can be negative) when $\delta_{a,k}$ and $\delta_{b,k}$ are chosen as 
\begin{equation}
\begin{aligned}
\delta_{a,k}\jw & = -[\sgn{\hat{a}_k\jw} ] \Delta_{a,k}\jw \\
\quad \delta_{b,k}\jw  & = -[\sgn{\hat{b}_k\jw}] \Delta_{b,k}\jw, 
 \end{aligned}
\end{equation}
which results in 
{\small{
\begin{equation}
\hat{g}_k\jw.g\jw     =  \hat{a}_k^2\jw +  \hat{b}_k^2\jw   -\Delta_{a,k}\jw|\hat{a}_k\jw|  -\Delta_{b,k}\jw|\hat{b}_k\jw|
\label{eq_P_k_condition_uncertainty_4}
\end{equation}
}}

\noindent
and the numerator in the right hand side  of Eq.~\eqref{eq_P_k_condition_uncertainty}.
The denominator in Eq.~\eqref{eq_P_k_condition_uncertainty_2}  is maximized when the system magnitude 
$| g\jw |$ is the largest possible, i.e., from  Eq.~\eqref{eq_a_delta_def}  
\begin{equation}
\begin{aligned}
\delta_{a,k}\jw & = [\sgn{\hat{a}_k\jw} ] \Delta_{a,k}\jw \\
\quad \delta_{b,k}\jw  & = [\sgn{\hat{b}_k\jw}] \Delta_{b,k}\jw, 
 \end{aligned}
\end{equation}
which results in the denominator in the right hand side  of Eq.~\eqref{eq_P_k_condition_uncertainty}.
Minimizing the numerator and maximizing the denominator of Eq.~\eqref{eq_P_k_condition_uncertainty_2} results in the  lemma's claim in 
Eq.~\eqref{eq_P_k_condition_uncertainty}. 

\hfill \qed

 \vspace{0.1in}
 \noindent
Next, an iterative control law is designed to reduce the tracking error  when the model uncertainties satisfy the bounds in Condition~\ref{condition_uncertainty_bounds}.

 \vspace{0.1in}
\begin{lemma}[Error reduction with iteration]
\label{lemma_IML_convergence_bounds}
Under Condition~\ref{condition_uncertainty_bounds}, 
and non-zero system model $\hat{g}_k\jw  \ne 0$  at frequency $\omega$, 
the iteration law in Eq.~\eqref{iteration_law_eq} reduces the tracking error, i.e., 
 \begin{equation}
| y_d\jw - y_k\jw |  < | y_d\jw - y_{k-1}\jw |
\label{eq_decreasing_error}
 \end{equation} 
provided the iteration gain $\rho_k$ satisfies 
 \begin{equation}
0 <  \rho_k < 
2 
\frac{ |\hat{g}_k\jw|^2  - \hat{g}_{abs,k}\jw.\Delta_k\jw}{
|\hat{g}_{abs,k}\jw  + \Delta_k\jw|^2
}  =  \rho_k^*\jw .
\label{eq_uncertianty_bound_k}
 \end{equation} 
 \end{lemma}
 
\vspace{0.01in}
\noindent
{\bf{Proof}:}
From Eqs.~\eqref{eq_M_k_condition_uncertainty}, ~\eqref{eq_P_k_condition_uncertainty}  and ~\eqref{eq_uncertianty_bound_k} 
 \begin{equation}
 0 <   \rho_k\jw  <  \rho_k^*\jw  \le  2  \frac{\cos[{\Delta_{p,k}}\jw] }{\Delta_{m,k}\jw}  . 
\label{eq_uncertianty_bound_proof_1}
 \end{equation}
 Since
$ {\Delta_{m,k}\jw} > 0$ (from Eq.~\eqref{eq_M_k_condition_uncertainty_derivation_1}, as the model $\hat{g}_k\jw$  and the system ${g}_k\jw$ are assumed to be non-zero at frequency $\omega$), 
    \begin{equation}
 \rho_k\jw  {\Delta_{m,k}\jw}    -2{\cos[{\Delta_{p,k}}\jw] }  < 0 
 \label{eq_uncertianty_bound_proof_5_new}
 \end{equation}
 and as $\rho_k\jw > 0$ from Eq.~\eqref{eq_uncertianty_bound_proof_1}, 
     \begin{equation}
    \begin{aligned}
 \rho_k\jw  {\Delta_{m,k}\jw}    \left[   \rho_k\jw  {\Delta_{m,k}\jw}    -2{\cos[{\Delta_{p,k}}\jw] }  \right]  & < 0  \\
 1 +  \rho_k\jw  {\Delta_{m,k}\jw}    \left[   \rho_k\jw  {\Delta_{m,k}\jw}    -2{\cos[{\Delta_{p,k}}\jw] }  \right]  &  < 1 .
   \end{aligned}
\label{eq_uncertianty_bound_proof_6_new}
 \end{equation}
 Then, 
 using re-arrangement of the terms similar to those in 
 Eqs.~\eqref{convergence_proof_2} and \eqref{convergence_proof_3}, results in 
  \begin{equation} 
\label{eq_uncertianty_bound_proof_7}
  \left|  1  -  \rho_k\jw  \frac{ g\jw }{ \hat{g}_k\jw }  \right|    < ~1
     \end{equation}
  and the lemma follows from  Eq.~\eqref{convergence_proof} with the iteration gain $\rho_k\jw$.
 \qed
 
 \vspace{0.1in}
\begin{rem}[Iteration gain]
\label{rem_iteration_gain_selection_rho}
 A large iteration gain can result in a large tracking-error reduction  but it can also 
amplify the effect of noise~\cite{szuchi_05}. Iteration gains larger than one can lead to oscillatory convergence. 
In the following, the iteration  gain $\rho_k\jw  $  is selected 
to satisfy  the upper bound $\rho_k^*\jw$ in Eq.~\eqref{eq_uncertianty_bound_proof_1}, as
\begin{equation}
\rho_k\jw = \underline{\rho}\jw  \min \left\{\rho_k^*\jw, 1\right\}  ~< \rho_k^*\jw, 
\label{eq_rho_k_selection}
\end{equation}
with  $\underline{\rho}\jw  \in (0,1)$.
 \end{rem}

 \vspace{0.1in}
\begin{rem}[Convergence with varying iteration gain]
\label{rem_iteration_gain_selection}
While the tracking error is decreasing as in Eq.~\eqref{eq_decreasing_error}, the rate of decrease can also potentially decrease 
with a varying iteration gain $\rho_k\jw$, e.g.,  $\rho_k\jw \rightarrow 0$. Nevertheless, if the modeling uncertainties decrease with additional data, $g_k\jw \rightarrow g\jw$, then 
the upper bound on the iteration gain increases $\rho_k^* \rightarrow 2$ from Eq.~\eqref{eq_uncertianty_bound_proof_1} and the iteration gain in Eq.~\eqref{eq_rho_k_selection} 
stays bounded away from zero, with $\rho_k\jw \rightarrow \underline{\rho}\jw $. Alternatively, 
the model updates and changes in the iteration gains could be stopped after a fixed number of iterations, say $k = k_{pe}$,  for guaranteed output-tracking convergence as in Lemma~\ref{lemma_convergence_single_input}
 \end{rem}

\vspace{0.1in}
\subsubsection{Zero iteration gain for large model uncertainty}
Guaranteed convergence of the iteratively found input  $u_k\jw$ to the desired 
inverse input $u_d\jw$ at frequency $\omega$, i.e.,  $u_k\jw \rightarrow u_d\jw$ in Eq.~\eqref{input_convergence_eq}, 
depends on the model uncertainties  $ \Delta_{a,k} \jw,  \Delta_{b,k} \jw $  being smaller than the size of the model
$ |\hat{a}_{k} \jw|,  |\hat{b}_{k}\jw| $. Therefore, the  upper bound on the iteration gain $ \rho_k^*\jw $  becomes  zero if the uncertainties are large, as stated in the lemma below.  

%
%
%

 \vspace{0.1in}
\begin{lemma}[Large uncertainty]
\label{lemma_acceptable_uncertainty_bounds}
If the uncertainties $\Delta_{m,k}\jw, \Delta_{p,k}\jw $ in the real $ \hat{a}_k\jw$ and imaginary $ \hat{b}_k\jw$ components of the model $\hat{g}_k$ are larger than the 
size of the model components, i.e., 
\begin{equation}
\label{model_uncertainty_unacceptable}
\begin{aligned}
\Delta_{a,k} \jw   &~~    \ge   | \hat{a}_k\jw |    \\
\Delta_{b,k} \jw       & ~~    \ge  |   \hat{b}_k\jw |,   \\
\end{aligned}
 \end{equation}
then, the upper bound $\rho_k^*\jw $  on the iteration gain  
in the convergence condition in Eq.~\eqref{eq_uncertianty_bound_k} of Lemma~\ref{lemma_IML_convergence_bounds}
 is zero
 \begin{equation}
\label{model_uncertainty_acceptable_zero}
\rho_k^*\jw  = 0.
 \end{equation} 
\end{lemma}
 
\vspace{0.01in}
\noindent
{\bf{Proof}:}
The lemma follows since,   the uncertainty $\Delta_k\jw = \hat{g}_{abs,k}\jw $, as defined in Eq.~\eqref{eq_P_k_condition_uncertainty_2}, satisfies  Eq.~\eqref{model_uncertainty_unacceptable} 
and results in 
\begin{equation} 
{ |\hat{g}_k\jw|^2  - \hat{g}_{abs,k}\jw.\Delta_k\jw} = 0.
\end{equation}
 \qed

\vspace{0.1in}
\begin{rem}[Magnitude and phase uncertainties]
\label{rem_magnitude_and_phase_uncertainties}
If the uncertainties $\Delta_{m,k}, \Delta_{p,k}$ in the real $ \hat{a}_k$ and imaginary $ \hat{b}_k$ components of the 
model $\hat{g}_k$ are larger than  the size of the model components, i.e.,   as in Eq.~\eqref{model_uncertainty_unacceptable}, 
the magnitude uncertainty $\Delta_{m,k}\jw$ can become infinite in Eq.~\eqref{eq_M_k_condition_uncertainty} 
and the cosine of the phase uncertainty $\cos[{\Delta_{p,k}}\jw] $  cannot be guaranteed to be greater than zero in Eq.~\eqref{eq_P_k_condition_uncertainty}, i.e.,
the phase uncertainty ${\Delta_{p,k}}\jw$ cannot be guaranteed to be less than $90^o$. 
These are required for guaranteed reduction in the output tracking error with iterations in previous results in~\cite{szuchi_05,hatonen_04}, 
e.g., as in Remark~\ref{rem_phase_error}. 
\end{rem}

\vspace{0.1in}
\begin{rem}[Similar conditions to robust inversion]
\label{robust_inversion_rem}
The requirement, that the  uncertainties $\Delta_{m,k}\jw, \Delta_{p,k}\jw $ in the model components  $ \hat{a}_k\jw, \hat{b}_k\jw$  should be  
smaller than the size of the  model components to ensure a nonzero iteration gain as  in Lemma~\ref{lemma_acceptable_uncertainty_bounds}, 
is similar to condition that the model uncertainty be smaller than the size of the model  for robust invertibility~\cite{sdrobust02}.
\end{rem}

\vspace{0.1in}
\begin{rem}[Limited tracking beyond system bandwidth]
\label{rem_limited_tracking_bw_wc}
At frequencies $\omega$ much higher than the system bandwidth $\omega_{bw}$, typical system model magnitudes $|\hat{g}\jw|$ 
tend to become small.  Therefore, for a given level of  model uncertainties 
$\Delta_{a,k} \jw, \Delta_{b,k} \jw$ (due to noise in the measurements), the 
upper bound  $ \rho^*\jw $ on the iteration gain in Eq.~\eqref{eq_uncertianty_bound_k} tends to 
be zero at  high frequencies $\omega >> \omega_{bw}$.
\end{rem}

\subsection{Iteration law with persistency of excitation}

Sufficient richness of the measured system-model  and reduced error in the model data  
can be achieved by augmenting the input ${u}_k\jw$ with an additional term  $\tilde{u}_{k}\jw$ that ensures persistency of excitation
at a frequency $\omega$ even if the desired output $y_d\jw$, and therefore the feedforward input $u_d\jw$ in Eq.~\eqref{inverse_system_eq} are zero at 
that frequency $\omega$ .

\vspace{0.1in}
\subsubsection{Modified iteration}
The iterative law in Eq.~\eqref{iteration_law_eq} is modified to 
\begin{equation}
\label{iteration_law_eq_noise}
\begin{aligned}
\hat{u}_{k}\jw ~& =  [u_{k-1}\jw  -\tilde{u}_{k-1}\jw ] \\
&  + \rho_{k}\jw  \hat{g}_k^{-1} \jw \left[   y_d\jw - ({y}_{m,k-1}\jw- \tilde{y}_{m,k-1}\jw) \right] \\
u_{k}\jw ~& =  \hat{u}_{k}\jw   + \tilde{u}_{k}\jw \\
\end{aligned}
\end{equation}
for $\omega \le \omega_c$ and zero elsewhere.  
Here the  additional input  $\tilde{u}_{k}\jw$ is used whenever 
the un-augmented 
input $\hat{u}_{k}\jw$ becomes small to provide persistence of excitation, in the first $k_{pe} < k^*$ iterations, where 
$k^*$ is the total number of iteration steps. 
For example, the additional input can be chosen to have 
magnitude $\tilde{u}\jw$ and a random phase angle 
$\phi_{k}$, i.e., 
\begin{equation}
\label{iteration_law_eq_unknoise}
\begin{aligned}
\tilde{u}_{k}\jw ~ & =
\left\{ 
\begin{aligned}
 &  \tilde{u}\jw  e^{j [\phi_{k}\jw] }  & \quad & {\mbox{if}}~| \hat{u}_{k}\jw | <  u_{pe}\jw  \\ 
 &                                                       &            &    {\mbox{and }} ~  k \le k_{pe}\\ 
& 0  &  \quad &   {\mbox{otherwise}}
\end{aligned}
 \right.  
\\
 \phi_{k}\jw  ~ &   \sim  {\mathcal{N}}\left(0,  \pi^2  \right), 
\end{aligned}
\end{equation}
where $u_{pe}\jw$ is selected to designate when the input is considered to be small. 
Note that the  additional input $\tilde{u}_{k-1}\jw$ from the previous iteration is removed when updating the iterative input 
$u_{k}\jw$ in Eq.~\eqref{iteration_law_eq_noise}. 
Moreover, the estimated effect $\tilde{y}_{m,k-1}\jw$ of the additional input $ \tilde{u}_{k-1}\jw$ on the measured output $y_{m,k-1}\jw$ 
is removed before computing the  updated input  
$u_{k-1}\jw$ in Eq.~\eqref{iteration_law_eq_noise}, with    
\begin{equation}
\label{iteration_law_no_output_noise}
\begin{aligned}
\tilde{y}_{m,k-1}\jw~ &  =  \hat{g}_k\jw  \tilde{u}_{k-1}\jw, 
\end{aligned}
\end{equation}
where the measured output $y_{m,k-1}\jw$  includes potential measurement noise $n_{y,k-1}\jw$, i.e., 
\begin{equation}
\begin{aligned}
\label{eq_output_measurement_noise}
y_{m,k-1}\jw ~ & = y_{k-1}\jw + n_{y,k-1}\jw  \\
n_{y,k-1}\jw  ~ &   \sim  {\mathcal{N}}\left(0,  \sigma_{y,a}^2\jw \right) + j {\mathcal{N}}\left(0,  \sigma_{y,b}^2\jw \right) .
\end{aligned}
\end{equation}
%

\vspace{0.1in}
\begin{rem}[Residual modeling error]
\label{rem_large_modeling_error}
If the model is not learned well, then the correction of the input augmentation in Eq.~\eqref{iteration_law_no_output_noise} will 
not be exact.  The error due to in-exact compensation of the input augmentation can be  corrected iteratively if 
the total number of iterations $k^*$ is sufficiently larger than the initial iterations $k \le k_{pe}$ when the input is augmented, i.e., $k_{pe} << k^*$. 
\end{rem}

\vspace{0.1in}
If an initial model $\hat{g}_1$ is available, then the output-tracking input can be estimated as  
\begin{equation}
\label{iteration_law_eq_noise_00}
\begin{aligned}
\hat{u}_{1}\jw ~& =   \hat{g}_1^{-1} \jw \left[   y_d\jw \right] ~~ \forall ~ \omega \le \omega_c, 
\end{aligned}
\end{equation}
and zero elsewhere. 
If an initial  model is not available, then $\hat{u}_1\jw=0$ for all frequency $\omega$. 
The input $\hat{u}_1$ can be augmented to improve model estimation, 
as 
\begin{equation}
\label{iteration_law_eq_noise_3}
\begin{aligned}
u_{1}\jw ~& =  \hat{u}_{1}\jw ~+\tilde{u}_{1}\jw, \quad 
~~ \forall ~ \omega \le \omega_c, 
\end{aligned}
\end{equation}
and zero elsewhere,  with $\tilde{u}_{1}$ as in Eq.~\eqref{iteration_law_eq_unknoise}.

%
%
%

\vspace{0.1in}
\begin{rem}[Initial input using learned model]
The final model  $\hat{g}_{k^*}$ at the final iteration step ${k^*}$ can be 
used as the initial model $\hat{g}_1$ in Eq.~\eqref{iteration_law_eq_noise_3} during the iterative output tracking of a new output 
trajectory $y_{d,2}$.
\end{rem}

\vspace{0.1in}
\subsubsection{Additional data for model}
New data $m_{a,k}\jw, m_{b,k}\jw$  to estimate the system model  $\hat{g}_k\jw$ can be computed
at frequencies $\omega$ where the input $u_{k-1}\jw$ is sufficiently large, say 
\begin{equation}
\label{model_estimation}
|u_{k-1}\jw|  ~ \ge  u_{ok}\jw > 0, 
\end{equation}
as in Eq.~\eqref{inverse_GPs_eq_2} 
\begin{equation}
\label{eq_additional_model_data}
\begin{aligned}
m_{g,k}\jw ~& = 
{m}_{a,k}\jw  + j {m}_{b,k}\jw  ~  = \frac{y_{m,k-1}\jw   }{u_{k-1}\jw} 
 \\ ~&  =
  \frac{y_{k-1}\jw  + n_{y, {k-1}}\jw }{u_{k-1}\jw} 
 \\ ~&  =
 g\jw ~
  +\frac{n_{y, {k-1}}\jw}{u_{k-1}\jw} 
\end{aligned}
\end{equation}
with the measurement noise $ n_{y,k-1}\jw$ as in Eq.~\eqref{eq_output_measurement_noise}.

\vspace{0.1in}
\begin{rem}[Reduction of measurement-noise effect]
The magnitude  $\tilde{u}\jw$ of the additional input at frequency $\omega$ in Eq.~\eqref{iteration_law_eq_unknoise}  should be selected to be much larger than the expected standard deviations
$\sigma_{y,a}\jw, \sigma_{y,b}\jw$  of the measurement noise $n_{y,k-1}\jw$ in Eq.~\eqref{eq_output_measurement_noise} 
to  increase the input $ u_{k-1}\jw$ and reduce the impact of measurement noise $n_{y,k-1}\jw$ on  the model data  ${m}_{a,k}\jw, {m}_{b,k}\jw$  computed using  Eq.~\eqref{eq_additional_model_data}. 
\end{rem}

\vspace{0.1in}
\subsubsection{Averaged model data}
The computational cost of GPR can become prohibitive as the number of model data increases, e.g., 
with increasing iteration steps. Therefore, in the proposed algorithm, the model data is first averaged at each frequency $\omega$ where $ N_{\omega,k}> 0$ data points are available similar to~\cite{Wang_zou_iterative_2015}, as discussed in Remark~\ref{rem_model_filtering}, 
\begin{equation}
\bar{m}_{g,k}\jw  ~ =  \bar{m}_{a,k}\jw  + j \bar{m}_{b,k}\jw  =  \frac{1}{N_{\omega,k}}  \sum_{i=1}^{k}  m_{g,i}\jw
\label{eq_averaged_data}
\end{equation} 
and $m_{g,i}$ is defined in Eq.~\eqref{eq_additional_model_data}. 
Then, the averaged data $\bar{m}_{g,k}\jw$ is used to  refine the model $\hat{g}_k$ using the GPR in Eq.~\eqref{general_GP_regression_mean_cov_eq} before computing the updated input $u_{k}\jw$ in Eq.~\eqref{iteration_law_eq_noise}. 
With sufficient number of data points $N_{\omega,k}$  at  a frequency $\omega \le \omega_c$ due to persistency of excitation,  
the  estimated mean $\bar{m}_{g,k}\jw$ of the model data  
tends to the system $g\jw$, e.g., as the number of iterations increase.


\subsection{Proposed algorithm}
\label{section_proposed_algorithm}
The proposed iterative machine learning (IML) algorithm using GPR is described below in Fig.~\ref{fig_algorithm}

\begin{figure}[htb]
\centering
  \begin{tabular}{@{}cc@{}}
    \includegraphics[width=.75\columnwidth]{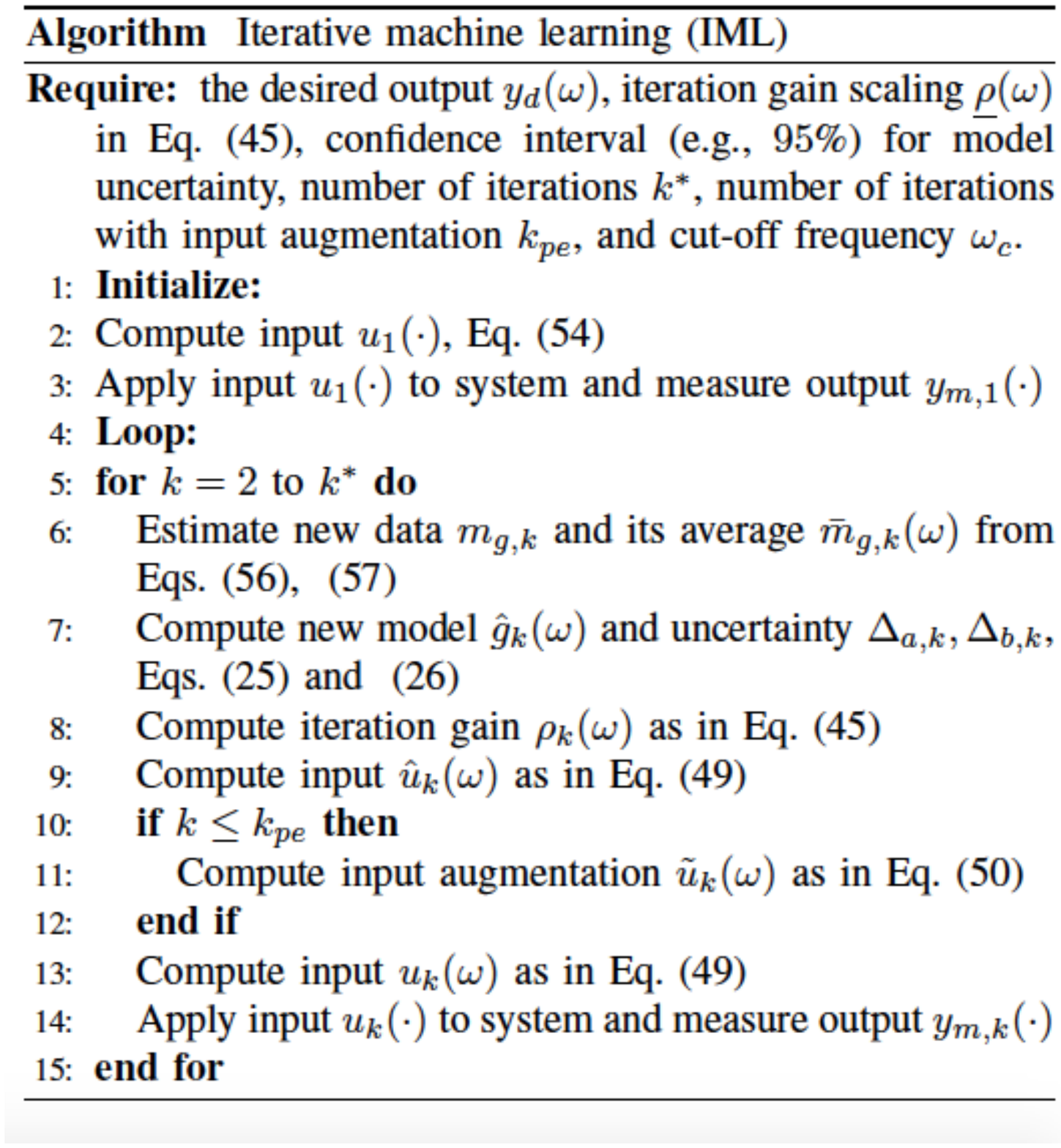} & 
  \end{tabular}
  \caption{{\em{ 
Iterative machine learning (IML)
}}}
\label{fig_algorithm}
\end{figure}


\vspace{0.1in}
\section{Simulation results and discussion}
\label{sec_results}

Simulation results are presented on convergence of the trajectory tracking input with the proposed IML approach. 
The impact of using the additional input for persistency of excitation on the learning of the model is illustrated. 
Additionally, the advantage of using the model learning to improve tracking of a new trajectory is illustrated. 

\vspace{0.1in}
\subsection{Example system}
\begin{figure}[htb]
\centering
  \begin{tabular}{@{}cc@{}}
    \includegraphics[width=.75\columnwidth]{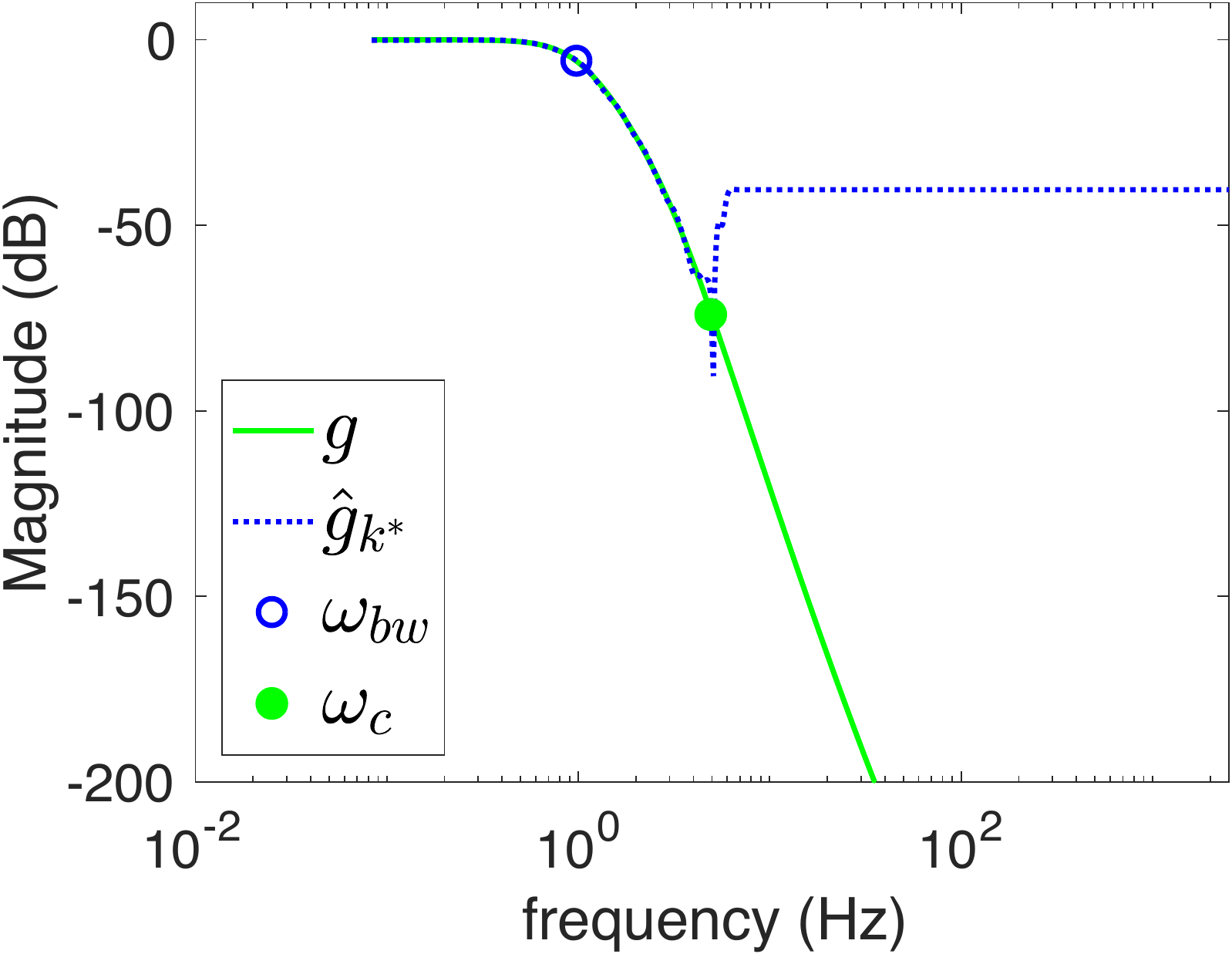} \\
    \includegraphics[width=.75\columnwidth]{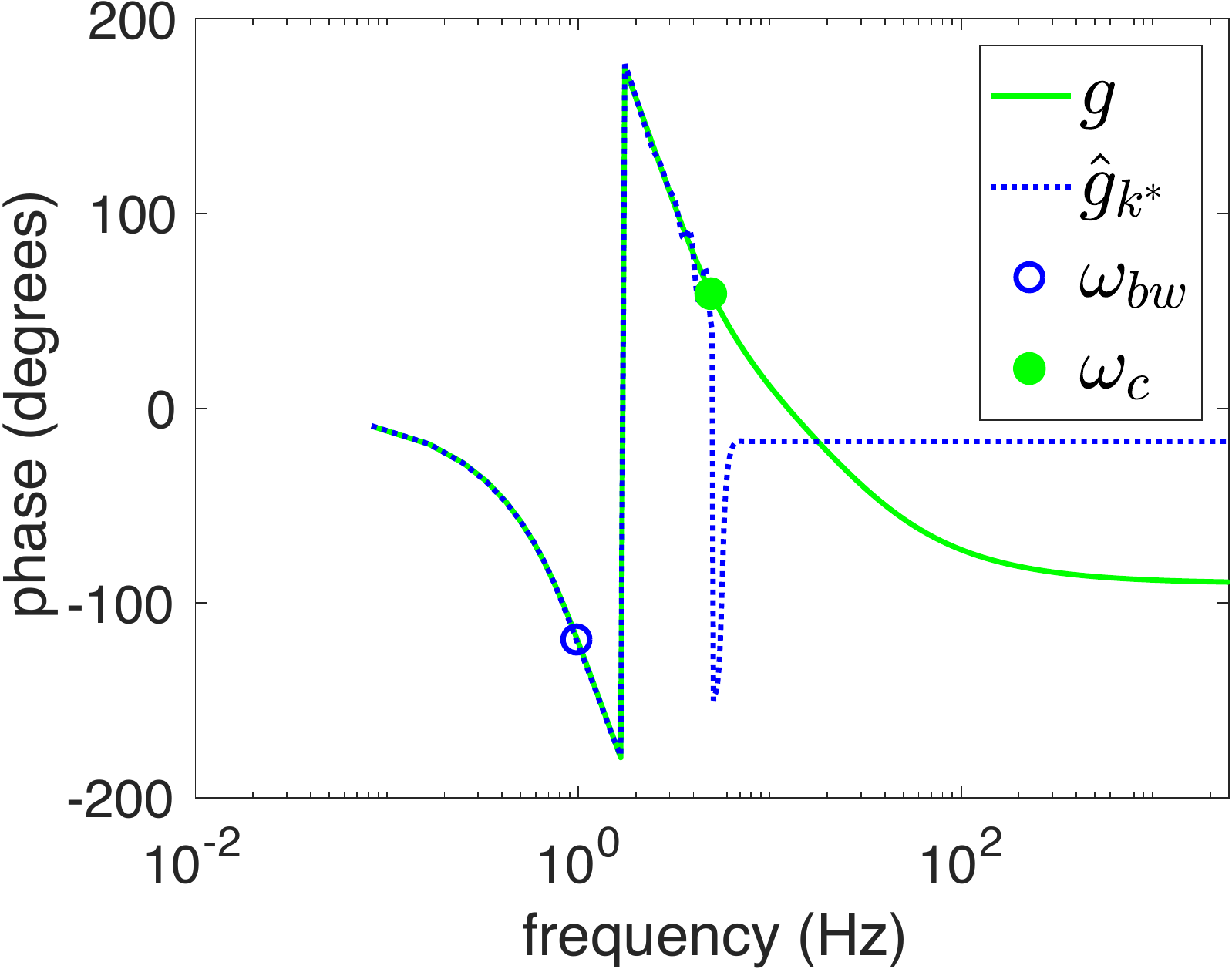} 
  \end{tabular}
  \caption{{\em{ 
Frequency response of the example system $g$ in Eq.~\eqref{E:GL}: (top) magnitude;  and (bottom)  phase  in degrees. 
The learned model $\hat{g}$ is also shown. 
}}}
\label{F1_system_freq_response}
\end{figure}

Consider a non-minimum phase   example system  $g(s)$ in Eq.~\eqref{system_eq} 
of the form 
\begin{equation} 
\label{E:GL}
g(s)  = 
\frac{  -\frac{\omega_{p1}^2 \omega_{p2}^2}{\omega_z^2}  
 \left[ (s -\omega_z) (s +\omega_z)\right] 
}
{
\left[ s^2 +2{\zeta_{p1}}{\omega}_{p1}s +\omega_{p1}^2 \right] 
\left[ s^2 +2{\zeta_{p2}}{\omega_{p2}}s +\omega_{p2}^2 \right]
} .
\end{equation} 
Let the larger pole frequency $\omega_{p2} = 6\pi$ rads/s be  
three times the smaller  pole frequency  $\omega_{p1}= 2\pi$ rads/s,  with the 
zero $\omega_{z} =4\pi$ rads/s  interlaced between the poles and the damping ratios  as 
$ \zeta_{p1} =  \zeta_{p2} = 1\sqrt{2}$. Note that the system has relative degree two (since the number of poles is two more than the number of zeros) 
and hence the desired output $y_d$ needs 
to be twice differentiable to enable exact tracking. 
The frequency response of the example system (with 
the above parameter values) is 
shown in Fig.~\ref{F1_system_freq_response}. The system bandwidth is about one Hz, $\omega_{bw} \approx 1$ Hz. 
Since output tracking is not usually anticipated much beyond the system bandwidth (see Remark~\ref{rem_limited_tracking_bw_wc}), the cutoff frequency $\omega_c$
for computing the model $\hat{g}$ and the iteration  input $u_k$ is selected as $\omega_c = 5$ Hz --- about five times the 
system  bandwidth $\omega_{bw}$.

\vspace{0.1in}
\subsection{Desired trajectory}
Consider a  twice-differentiable, desired trajectory $y_d$, with zero initial position $y_d(0)=0$ and velocity $\dot{y}_d(0)=0$, 
and  specified by its second time derivative  $\ddot{y}_d$ as 
\begin{equation}
\label{desired_output_eq}
\ddot{y}_d (t)  = 
\left\{ 
\begin{aligned}
 & \sum_{n=1}^{n=N} \sin [n \omega_* (t -t_0)  ]  & \quad & {\mbox{if}}~ t_0  < ~ t < t_1   \\ 
& - \sum_{n=1}^{n=N} \sin [n \omega_* (t -t_1)  ]    & \quad & {\mbox{if}}~ t_1  < ~ t < t_2   \\ 
& 0 & \quad &    {\mbox{otherwise}}, 
\end{aligned}
 \right.  
\end{equation}
where  main frequency component $\omega_* =  0.5$ Hz that is about half the system bandwidth $\omega_{bw}$,  
the number of harmonics $N$ in the desired output $y_d$ was one, i.e.,   $N=1$ 
and  
$t_0 = 4$s, $t_1 = 6$s, $t_2 = 8$s.
A plot of the desired output is shown in Fig.~\ref{F3_impact_u_n}.
The padding around the middle section $t\in[4,8]$ is added to enable noncausal solutions. 
The computations are performed in the time domain with a sampling time of $0.2$ms and  in the discrete-frequency domain with the fast Fourier transform (FFT) using MATLAB.

\vspace{0.1in}
\subsection{Impact of additional input for persistency of excitation}
The error in the model data from Eq.~\eqref{eq_additional_model_data} can be large in the presence of noisy output measurements $y_{m,k}\jw$. 
To illustrate, the error in the model data are compared below, with and without the persistency of excitation,  when the 
input  $u\jw$ is the exact tracking input $u_d\jw$  from Eq.~\eqref{inverse_system_eq}, for 
frequency $\omega$ less than the cutoff frequency $\omega_c$.

\begin{figure}[htb]
\centering
  \begin{tabular}{@{}cc@{}}
      \includegraphics[width=.45\columnwidth]{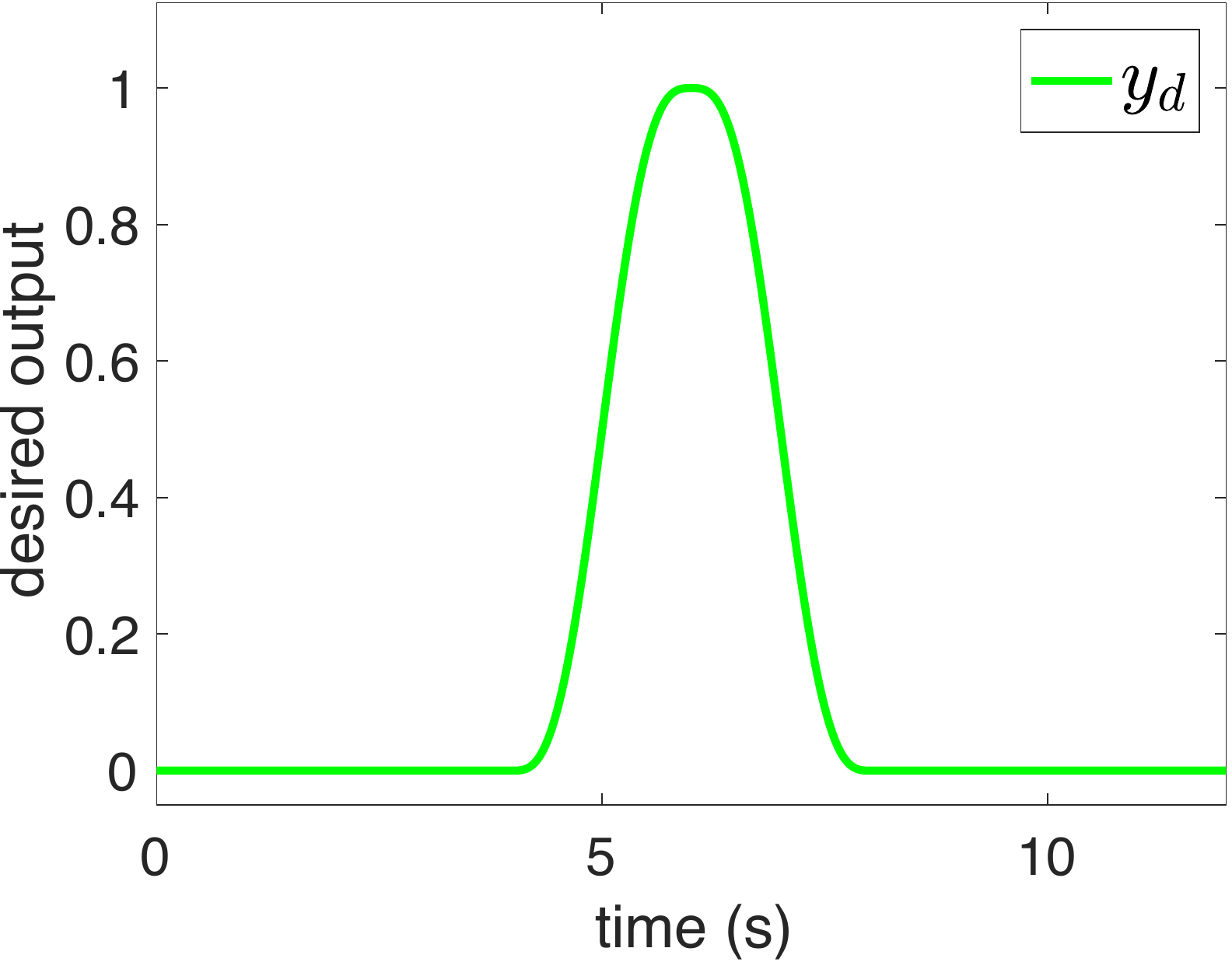} & 
    \includegraphics[width=.45\columnwidth]{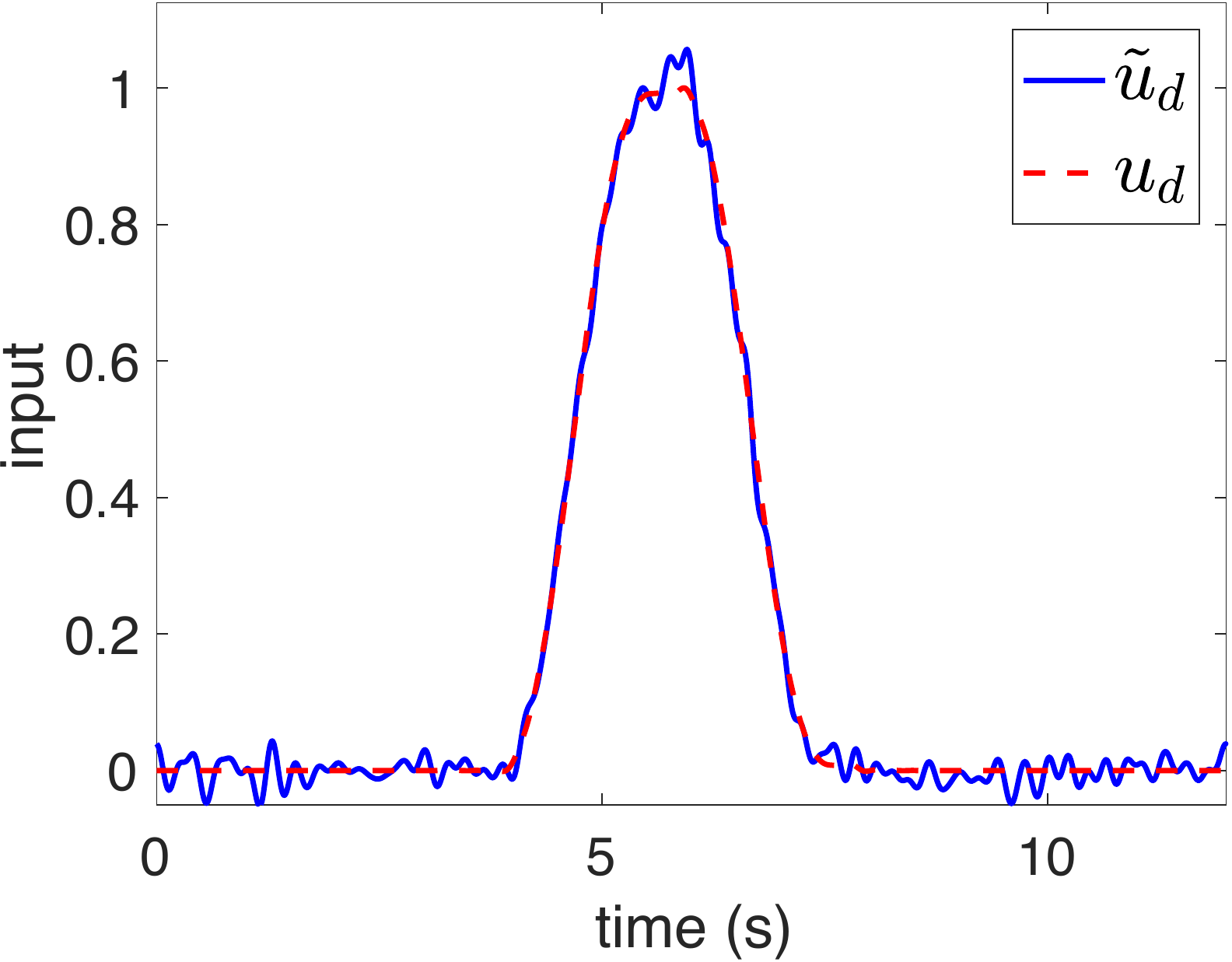} \\
    \includegraphics[width=.45\columnwidth]{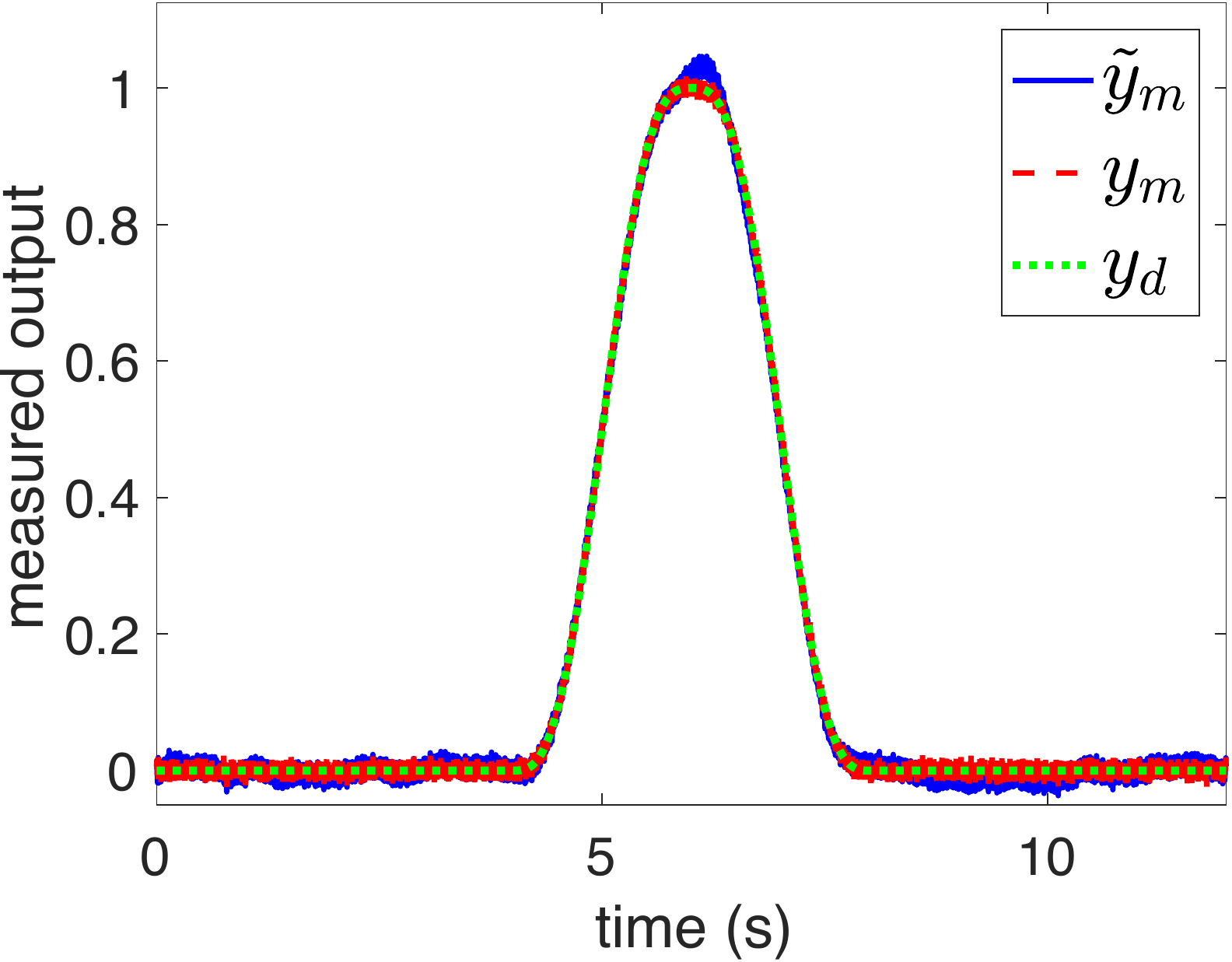} & 
\includegraphics[width=.45\columnwidth]{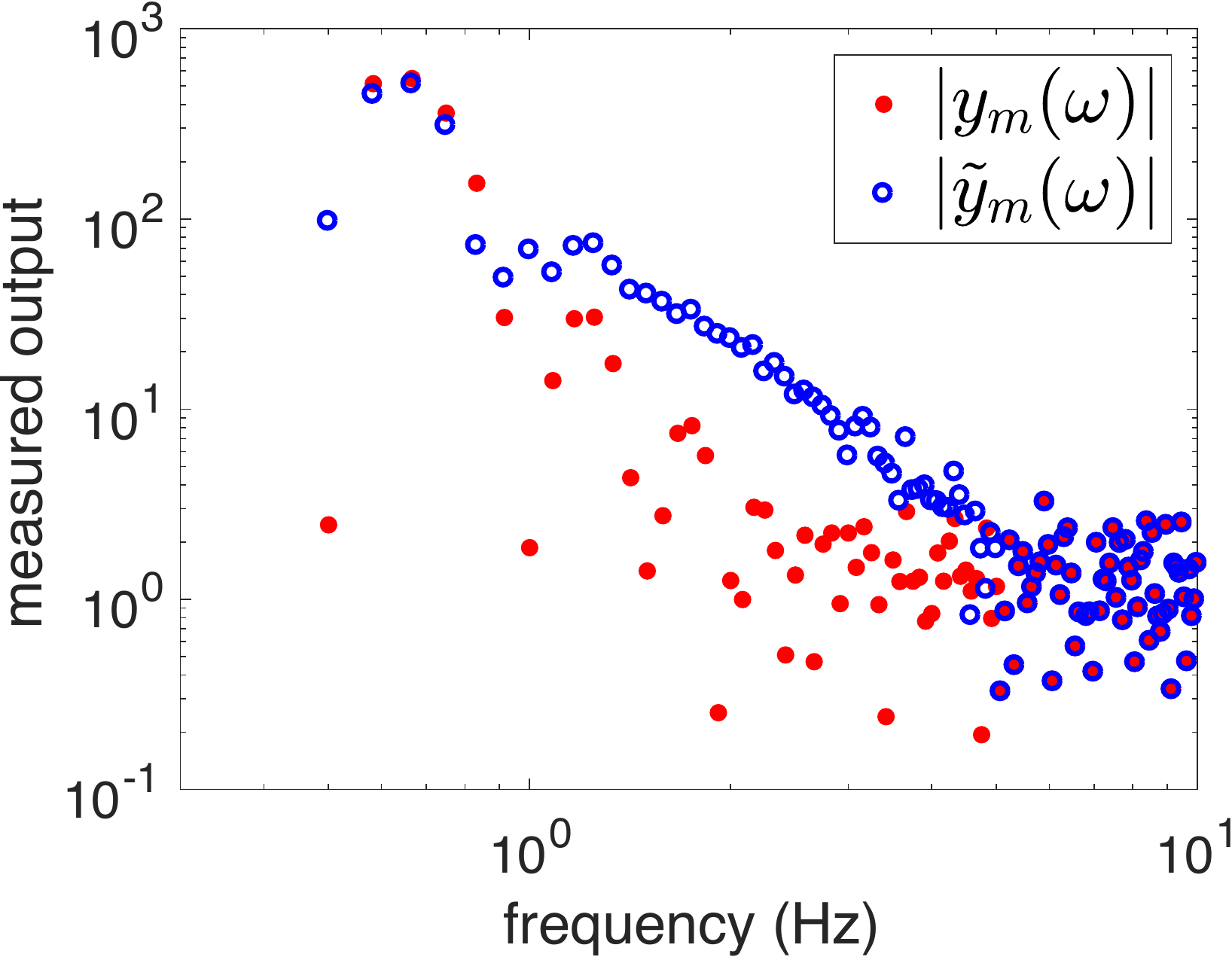}  
  \end{tabular}
  \caption{{\em{ 
Comparison of input and output with and without persistency of excitation. Desired output (top left). 
Input $u$ (top right) : (case i)~the inverse input $u_d$ for the desired output $y_d$  from Eq.~\eqref{inverse_system_eq}; 
and  (case ii)~the augmented input $\tilde{u}_{d}$ from Eq.~\eqref{iteration_law_eq_unknoise_ud} with persistency of excitation.
Resulting measured output $y_m$ from Eq.~\eqref{eq_output_measurement_noise_ud} and $\tilde{y}_m$ from Eq.~\eqref{eq_output_measurement_noise_ud_tilde} 
are shown in time domain (bottom left) and frequency domain (bottom right).  The plot of the measured outputs $y_m, \tilde{y}_m$ (bottom left) overlaps the plot of the desired output $y_d$.}}}
\label{F3_impact_u_n}
\end{figure}

 The standard deviation $\sigma_{n_y}\jw$ of the output noise 
$n_{y,d}\jw$  in the simulations is, similar to  ~\eqref{eq_output_measurement_noise}, 
\begin{equation}
\label{bounds_noise_example}
\sigma_{n_y} ~~= \sigma_{y,a}\jw ~=  \sigma_{y,b}\jw = \frac{1}{10^{4}}   \max_{\omega  \le \omega_c}  \left[ {y}_d\jw \right]
\end{equation}
and the resulting input and output are shown in Fig.~\ref{F3_impact_u_n}.

\vspace{0.1in}
\subsubsection{Effect of measurement noise}
\begin{figure}[htb]
\centering
  \begin{tabular}{@{}cc@{}}
    \includegraphics[width=.49\columnwidth]{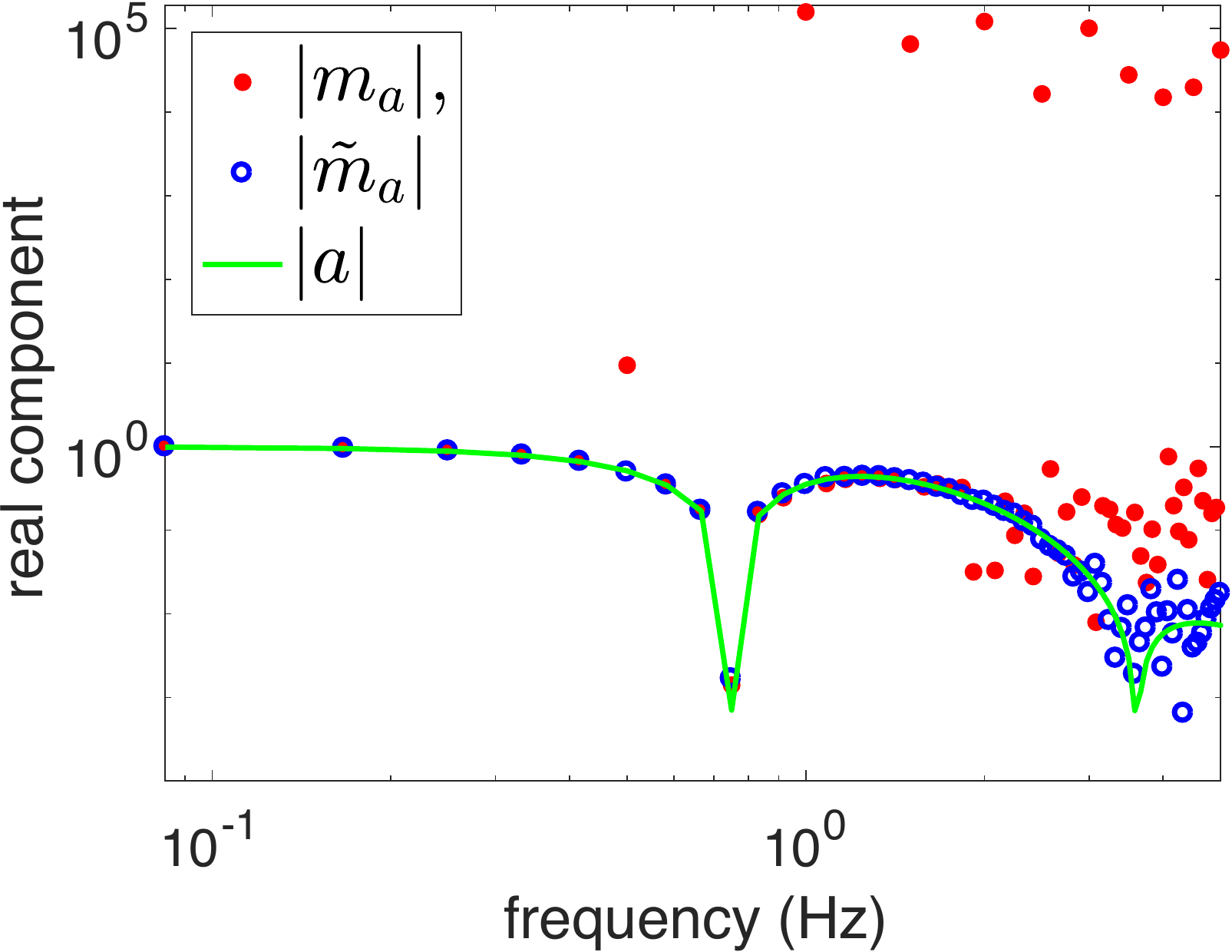} & 
              \includegraphics[width=.49\columnwidth]{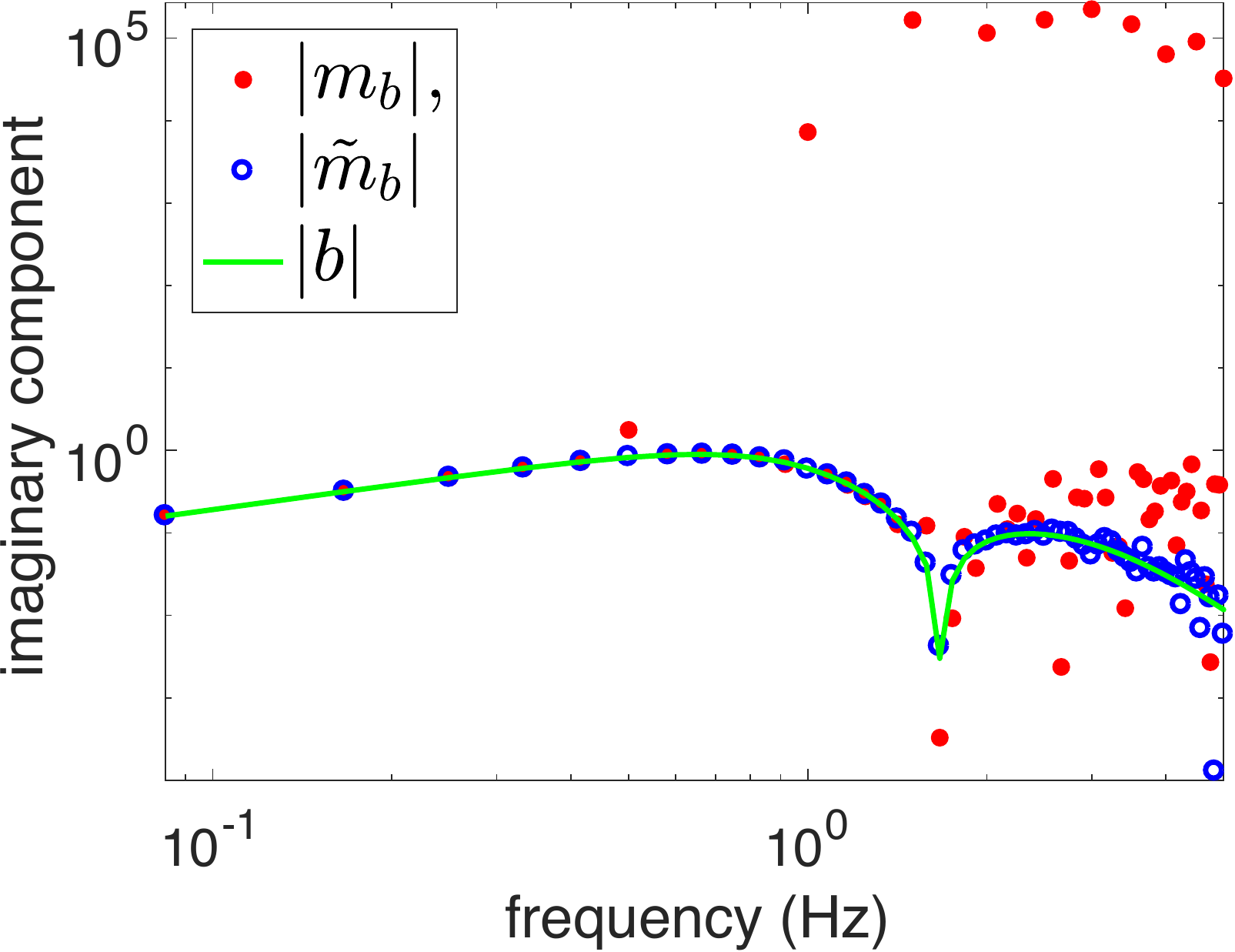}\\
     \includegraphics[width=.49\columnwidth]{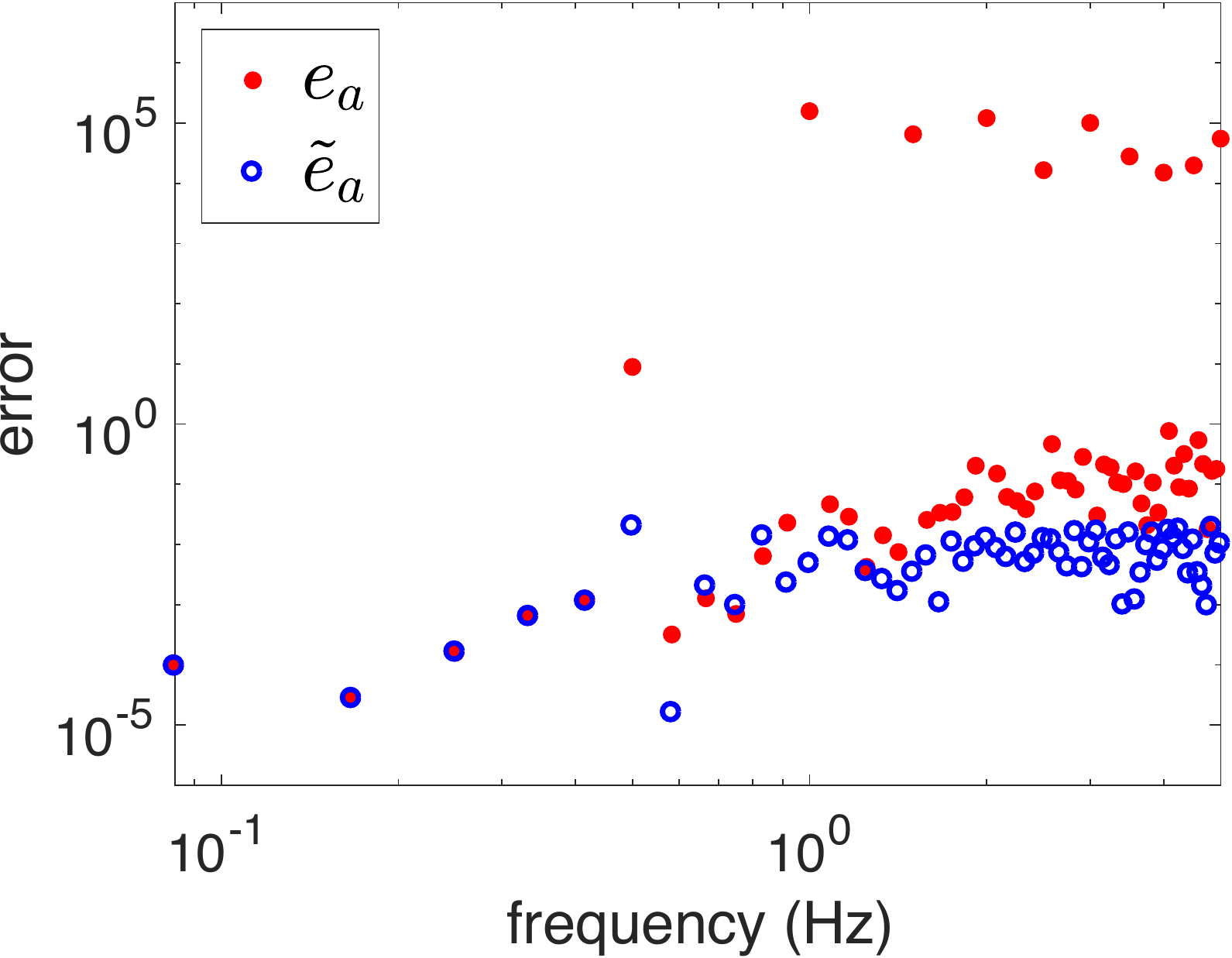} & 
    \includegraphics[width=.49\columnwidth]{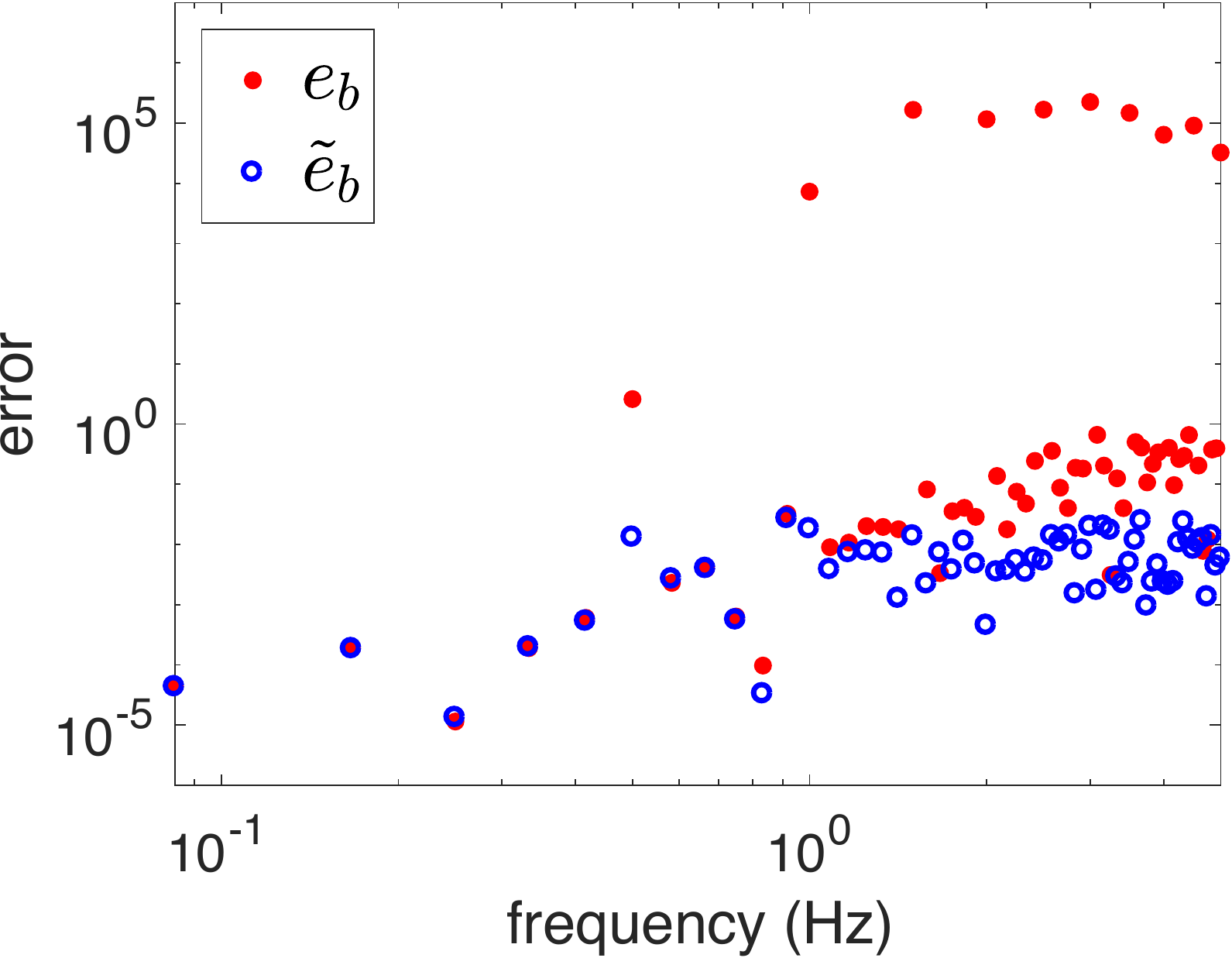} 
  \end{tabular}
  \caption{{\em{ 
Impact of persistency of excitation on acquired model data. 
Top two plots compare the absolute values of the real and imaginary components $a, b$ for the  
example system $g$ in Eq.~\eqref{E:GL} with the   components of the data $m_a, m_b$ from Eq.~\eqref{eq_additional_model_data_2} and 
$\tilde{m}_a, \tilde{m}_b$ from Eq.~\eqref{eq_additional_model_data_3}.
The bottom two plots compare the model-data error for the two cases, $e_a, e_b$ in  Eq.~\eqref{error_eab} and $\tilde{e}_a, \tilde{e}_b$ in  Eq.~\eqref{error_eab_tilde}.
}}}
\label{F4_error_model}
\end{figure}

The noisy measured 
output $y_{m}$, as in  Eq.~\eqref{eq_output_measurement_noise}, 
\begin{equation}
\label{eq_output_measurement_noise_ud}
y_{m,d}\jw ~  = y_{d}\jw + n_{y,d}\jw  ~ =  g\jw {u}_d\jw + n_{y,d}\jw 
\end{equation}
shown in Fig.~\ref{F3_impact_u_n} 
is close to the desired output $y_d$  with the inverse feedforward input $u=u_d$.
With this measured output $y_{m,d}$, the model data can be computed from Eq.~\eqref{eq_additional_model_data} as 
\begin{equation}
\label{eq_additional_model_data_2}
m_g\jw ~~ = {m}_{a}\jw  + j {m}_{b}\jw  ~ = \frac{y_{m,d}\jw}{u_d\jw}  .
\end{equation}
%
The error $e_a, e_b$ in the model data  components $m_a, m_b$ 
increases with frequency since  the output $y_d$  tends to become  small compared to the output noise $\sigma_{n_y}$, e.g., at frequencies, $\omega >1$ Hz,  as seen in Fig.~\ref{F4_error_model}~(bottom two plots). 
The maximum model-data error $ e_{a,max},  e_{b,max}$ in the real and imaginary components of the model data are  
\begin{equation}
\label{error_eab}
\begin{aligned}
  e_{a,max} ~ & =  \max_{\omega  \le \omega_c}  \left[ e_a\jw =  { | m_a\jw-a\jw |} \right]  ~ = 1.58 \times 10^5 \\
   e_{b,max} ~ & =  \max_{\omega  \le \omega_c}  \left[ e_b\jw = { | m_b\jw-b\jw |}  \right] ~ = 2.24 \times 10^5.
\end{aligned}
\end{equation}

\vspace{0.1in}
\subsubsection{Reduction of noise effect with persistency of excitation}
To evaluate the advantage of persistency of excitation,  consider the augmentation of the inverse input ${u}_d\jw$ to  $\tilde{u}_{d}\jw$, 
as in Eq~\eqref{iteration_law_eq_unknoise}, at frequency $\omega$ less than the cutoff frequency $\omega_c$, 
\begin{equation}
\label{iteration_law_eq_unknoise_ud}
\tilde{u}_{d}\jw ~  =  u_d\jw + \tilde{u}\jw  e^{j [\phi_d\jw] }  \\
\end{equation}
whenever the input $u_d\jw$ was  small, 
\begin{equation}
|{u}_d\jw| < u_{pe}\jw = 10\sigma_{n_y}
\end{equation}
with the additional input magnitude and phase given by 
\begin{equation}
\tilde{u}\jw ~~= 100 \sigma_{n_y}, \quad  
 \phi_d\jw  ~   \sim  {\mathcal{N}}\left(0,  \pi^2  \right) , 
\quad \omega \le \omega_{c}.
\end{equation}
The  resulting augmented input $\tilde{u}_{d}$ and the associated measured output $\tilde{y}_{m,d}$ with the same noise $n_{y,d}\jw$ as the unaugmented case 
in Eq.~\eqref{eq_output_measurement_noise_ud}
\begin{equation}
\label{eq_output_measurement_noise_ud_tilde}
\tilde{y}_{m,d}\jw ~  = g\jw \tilde{u}_d\jw + n_{y,d}\jw 
\end{equation}
are shown in Fig.~\ref{F3_impact_u_n}. 
The effect of measurement noise in the model estimates  from Eq.~\eqref{eq_additional_model_data} 
\begin{equation}
\label{eq_additional_model_data_3}
\tilde{m}_g\jw ~~ = \tilde{m}_{a}\jw  + j \tilde{m}_{b}\jw  ~ = \frac{\tilde{y}_{m,d}\jw}{\tilde{u}_d\jw} .
\end{equation}
%
%
are shown in Fig.~\ref{F4_error_model} (top two plots), which compares the 
 real and imaginary components $a\jw, b\jw$ of the example system $g\jw$ in Eq.~\eqref{E:GL}
 with the    real and imaginary  components of the model data  estimated for the two cases: 
(case i)~the inverse input $u_d\jw$ from Eq.~\eqref{inverse_system_eq} 
and  (case ii)~the input $\tilde{u}_{d}\jw$ with persistency of excitation from Eq.~\eqref{iteration_law_eq_unknoise_ud}. 
Note that the model data with the persistency~of~excitation input $\tilde{u}_{d}\jw$ tend to be closer to the actual system $g\jw$
when compared to the case without the 
persistency of excitation as seen in Fig.~\ref{F4_error_model}. 
The  maximum model-data error $ \tilde{e}_{a,max},  \tilde{e}_{b,max}$ in the real and imaginary components, with the persistency of excitation input, 
over all frequencies $\omega \le \omega_c$,  are  
\begin{equation}
\label{error_eab_tilde}
\begin{aligned}
 \tilde{e}_{a,max}   ~ & =  \max_{\omega  \le \omega_c} ~ \left[  \tilde{e}_a\jw =  {| \tilde{m}_a\jw-a\jw |} \right] ~ = 0.020,  \\
   \tilde{e}_{b,max} ~ & =   \max_{\omega  \le \omega_c} ~ \left[ \tilde{e}_b\jw = {| \tilde{m}_b\jw-b\jw |}  \right]~ = 0.027. 
   \end{aligned}
\end{equation}
Thus, the addition the  persistency of excitation input (with relatively-small change on the output, typically at high-frequency, as seen in Fig.~\ref{F3_impact_u_n}) 
can lead to  substantially smaller model error 
$ \tilde{e}_{a,max},  \tilde{e}_{b,max}$  in Eq.~\eqref{error_eab_tilde}
than the model error $e_{a,max}, e_{b,max}$ in Eq.~\eqref{error_eab} without the persistency of excitation --- several orders of magnitude less error, as seen in Fig.~\ref{F4_error_model}, (bottom two plots). 

\vspace{0.1in}
\subsection{Convergence with iterations}
Convergence with the proposed IML approach is discussed below with the 
initial input $u_1$ chosen with the augmented input in Eq.~\eqref{iteration_law_eq_noise_3} and without prior knowledge of the model, i.e., 
$\hat{u}_{1}\jw=0$ in Eq.~\eqref{iteration_law_eq_noise_3}. 

\vspace{0.1in}
\subsubsection{GPR results}
\begin{figure}[htb]
\centering
  \begin{tabular}{@{}cc@{}}
    \includegraphics[width=.49\columnwidth]{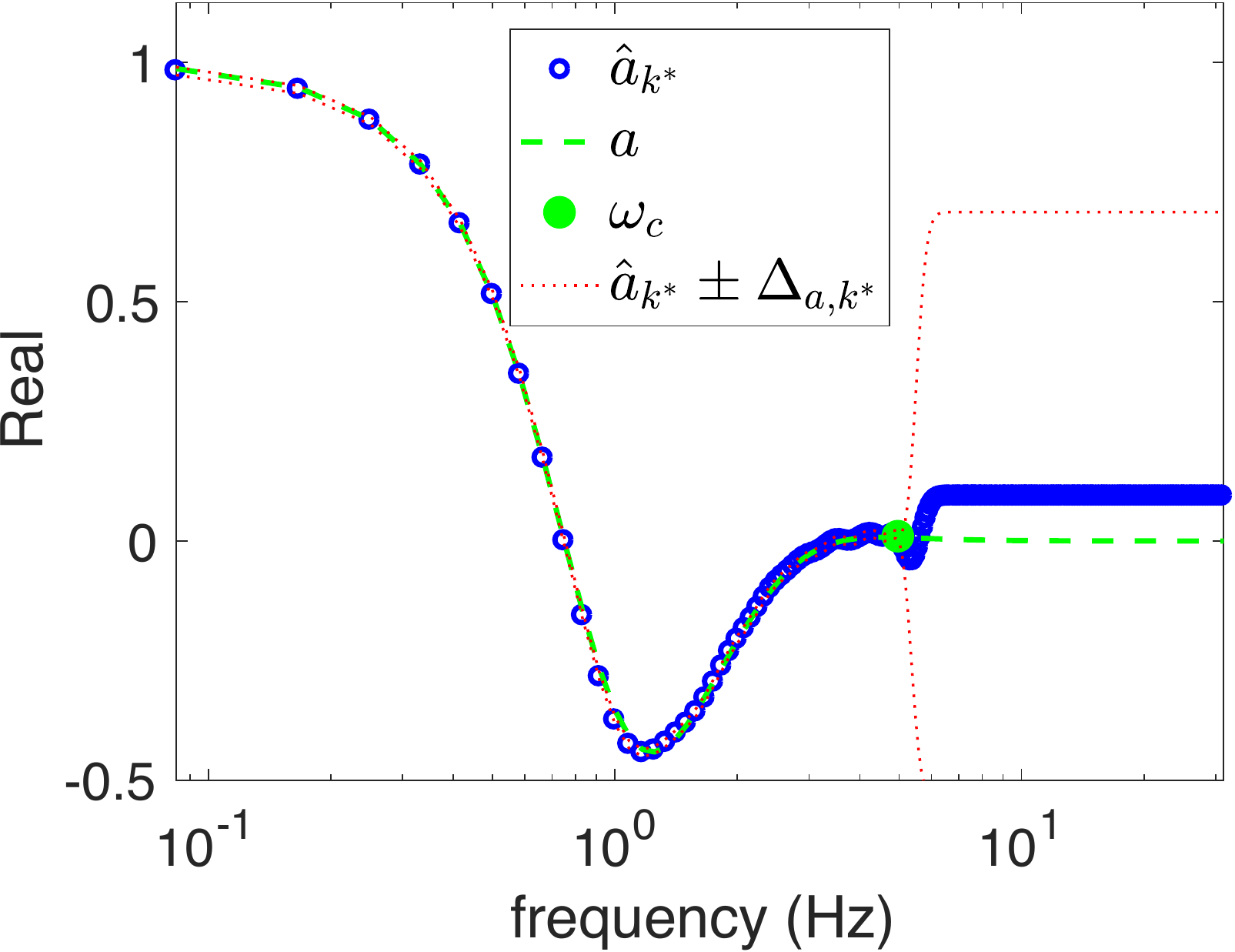} & 
    \includegraphics[width=.49\columnwidth]{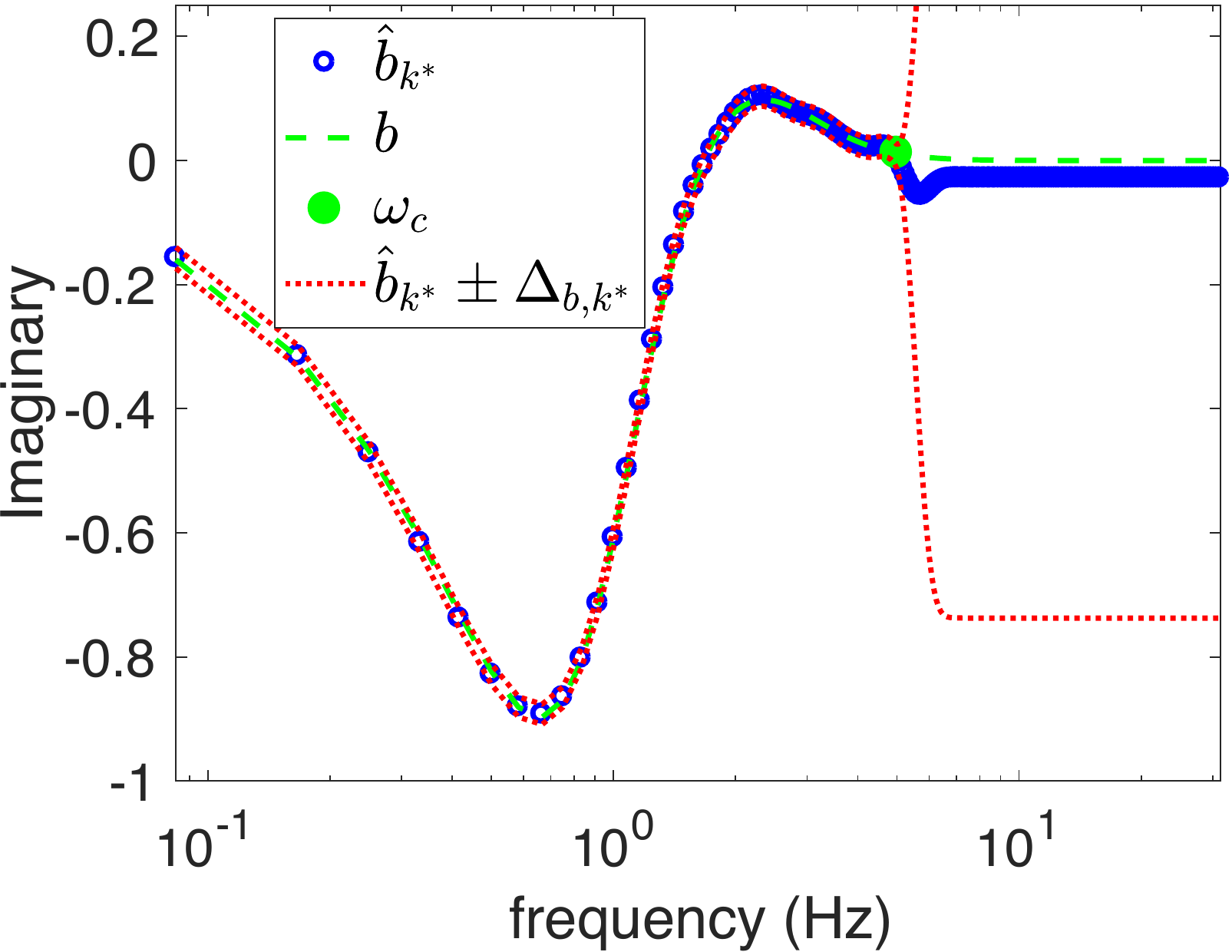} 
  \end{tabular}
  \caption{{\em{ 
Results of the Gaussian process regression. 
 The fit from the GPR 
 $\hat{a}_{k^*}\jw,  \hat{b}_{k^*}\jw$ is close to the system  ${a}\jw,  {b}\jw$  --- the fit overlaps the model upto the cutoff frequency $\omega_c$ due to the augmented input 
 and deviations become noticeable beyond $\omega_c$. 
The 95\% confidence interval (dotted lines) is small  till  the cutoff frequency $\omega_c$. 
}}}
\label{F4_modeling_GRP_data}
\end{figure}

All available data  ${m}_{g,i}\jw$  with $ i = 1, 2, \hdots k$  is used to predict the model $\hat{g}_k$ using GPR with 
{\em{fitrgp}} and  {\em{predict}} functions in MATLAB, where the covariance function is the squared exponential kernel, and the hyperparameters are optimized using the data. 
With the total number of iterations selected as  $k^*=5$ and  model augmentation during the first three iterations $k_{pe} = 3$ in Eq.~\eqref{iteration_law_eq_unknoise}, 
the  estimated 
the model data  $\hat{g}_{k^*}\jw$
is close to the system $g\jw$  as seen in Fig.~\ref{F4_modeling_GRP_data}.
The $95\%$ confidence intervals $\Delta_{a,k}, \Delta_{b,k}$ (e.g.,  shown in Fig.~\ref{F4_modeling_GRP_data} for the final iteration $k=k^*$) are used as the expected bounds on the model uncertainty as in Eq.~\eqref{model_uncertainty_ri_CI}.  

\vspace{0.1in}
\subsubsection{Selection of iteration gain}
\begin{figure}[htb]
\centering
  \begin{tabular}{@{}c@{}}
    \includegraphics[width=.6\columnwidth]{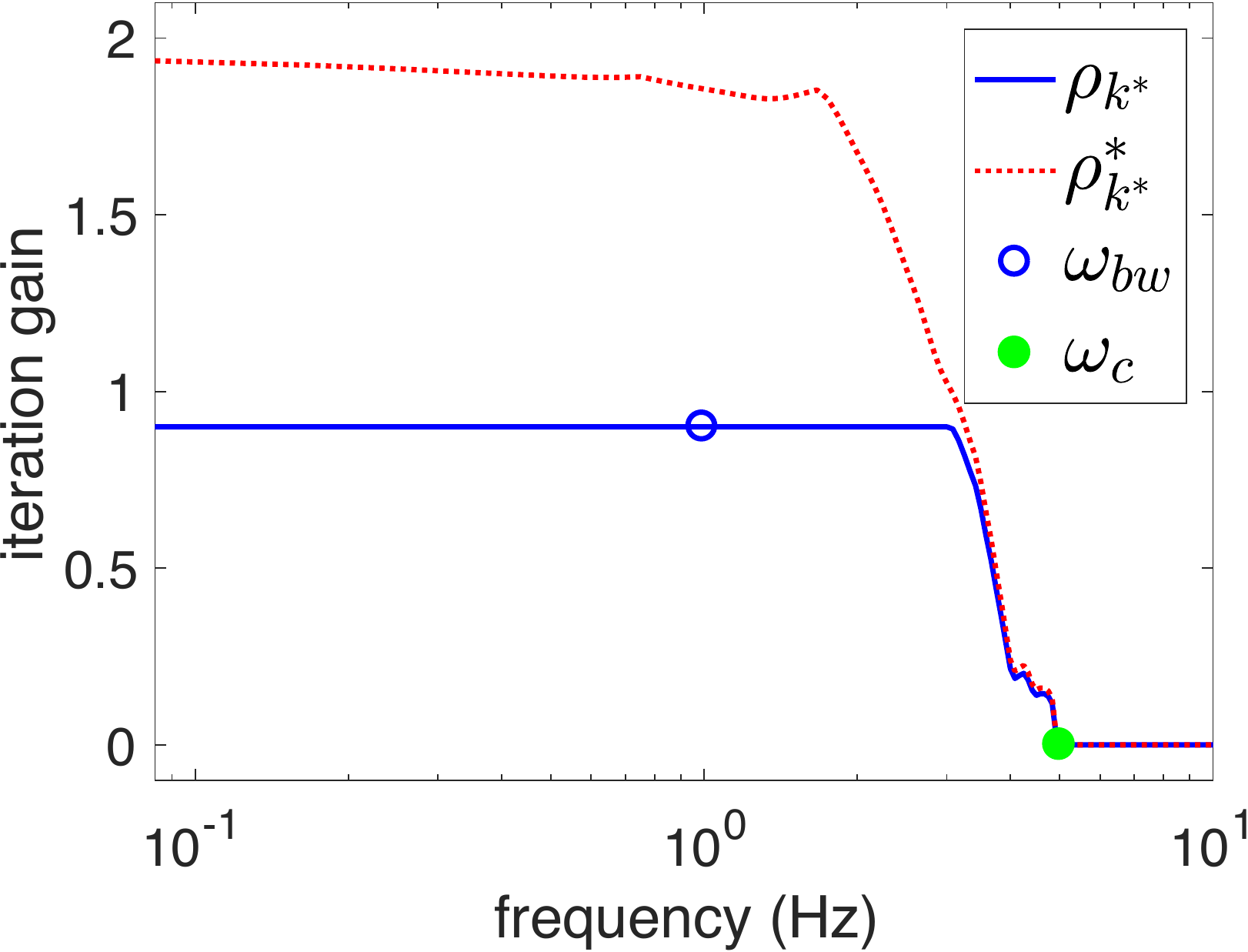} 
  \end{tabular}
  \caption{{\em{ 
The maximum possible iteration gain $\rho_{k^*}^*\jw$ from Eq.~\eqref{eq_uncertianty_bound_proof_1} and the selected iteration gain 
 $\rho_{k^*}$ used to compute the final input $u_{k^*}\jw$.
}}}
\label{F_6_iteration_gain}
\end{figure}

The  iteration gain $\rho_k\jw$ was chosen as in Eq.~\eqref{eq_rho_k_selection} with $ \underline{\rho}\jw  = 0.9$.   
The iteration gain  $\rho_{k^*}\jw$ and its upper bound  $\rho_{k^*}^*\jw$ from Eq.~\eqref{eq_uncertianty_bound_proof_1}  at the final iteration step $k^*$ are shown in Fig.~\ref{F_6_iteration_gain}, which tends to zero at high frequency when the model size becomes small compared to the measurement noise, as in Remark~\ref{rem_limited_tracking_bw_wc}.

\vspace{0.1in}
\begin{rem}[Cutoff frequency]
\label{rem_Cutoff_frequency}
While the iteration gain $\rho\jw$ acts as a low-pass filter and tends to zero at high frequency $\omega$ when the system $g\jw$ tends to zero, e.g., see Fig.~\ref{F_6_iteration_gain}, 
the 
cutoff frequency $\omega_c$ (beyond which the iteration gain $\rho\jw$ is set to zero) is used to avoid potential divergence at high-frequencies where the model size and the 
allowable uncertainty are expected to be small. 
\end{rem}

\vspace{0.1in}
\subsubsection{Input and model convergence}
\begin{figure}[htb]
\centering
  \begin{tabular}{@{}cc@{}}
  \includegraphics[width=.45\columnwidth]{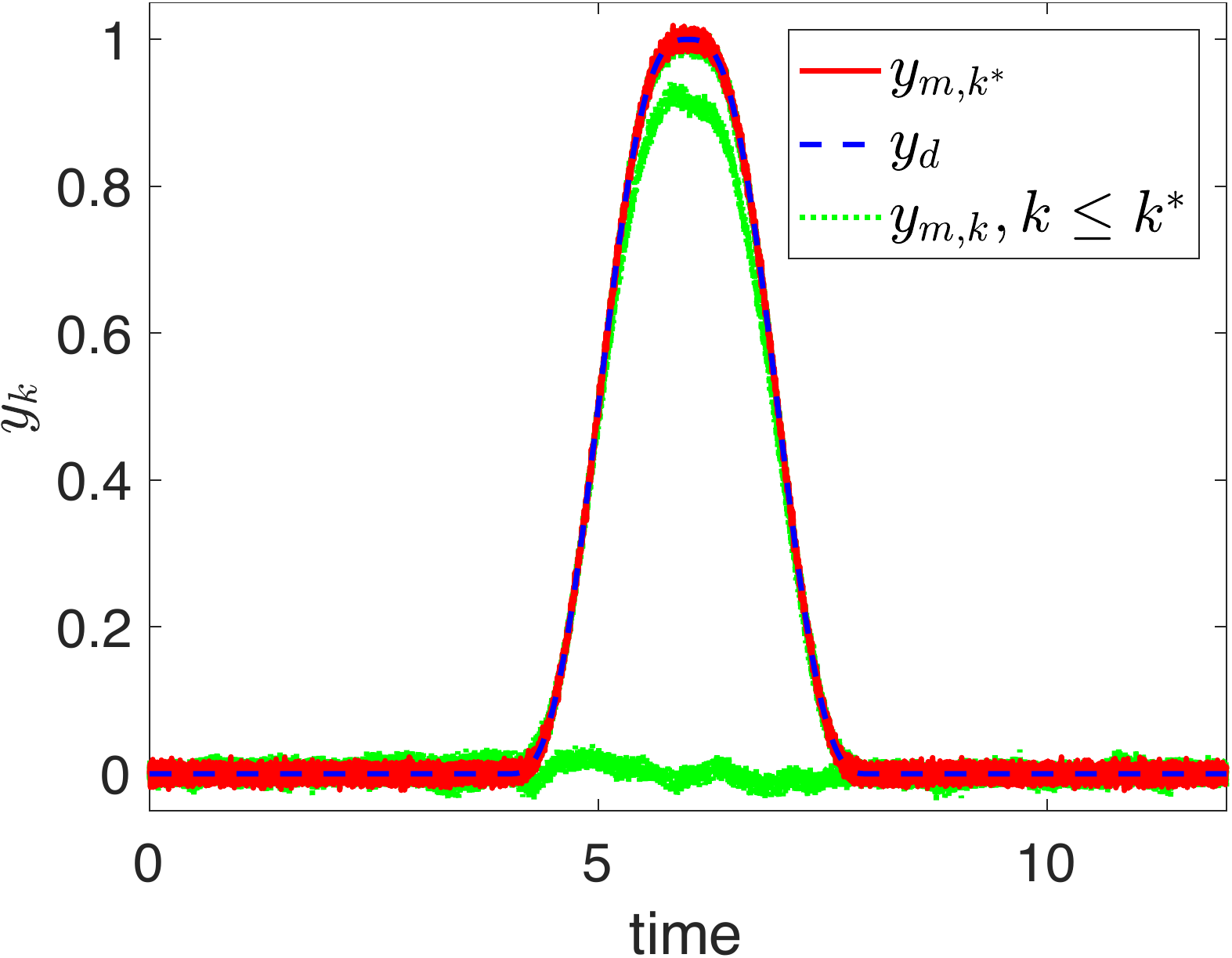} & 
    \includegraphics[width=.45\columnwidth]{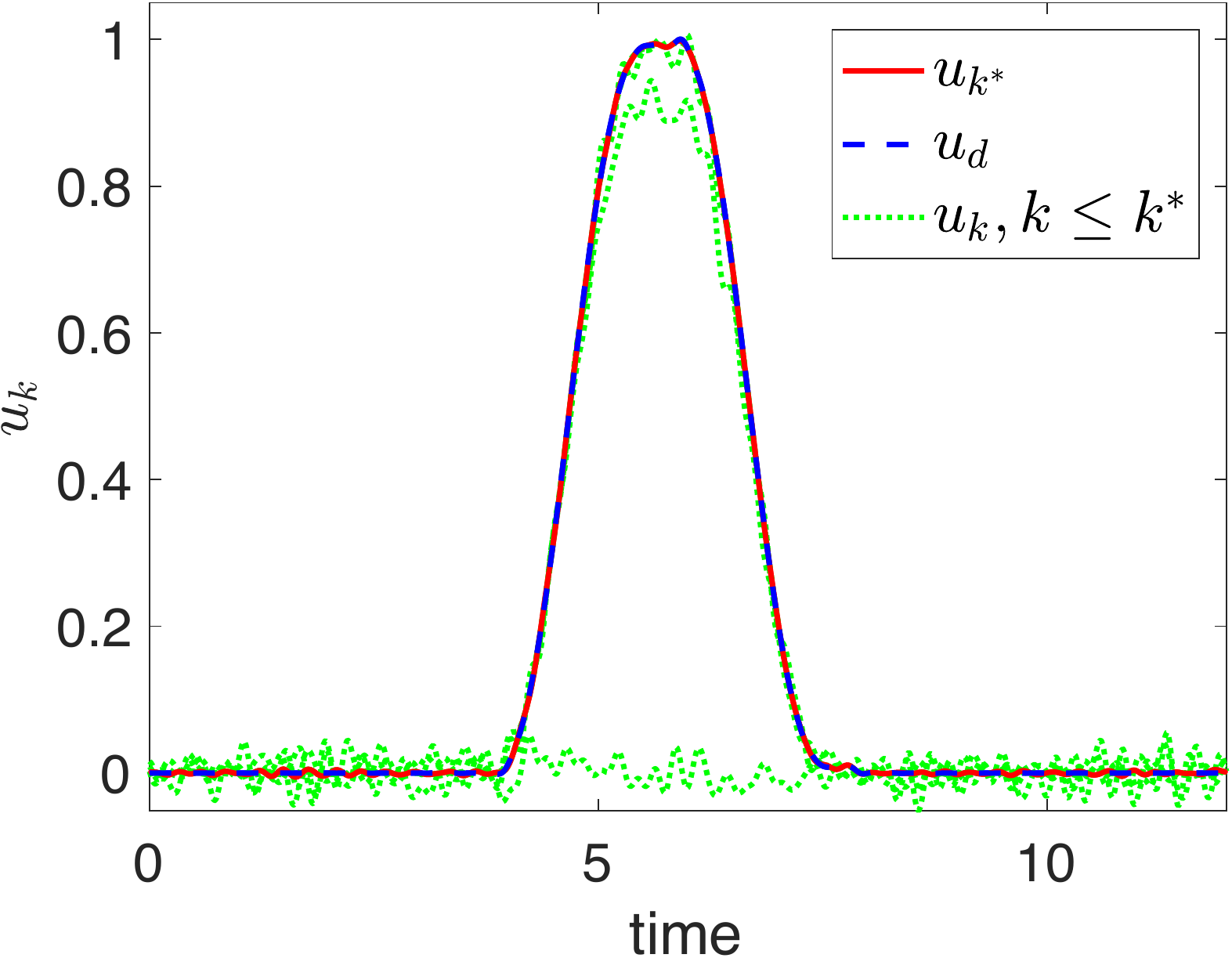} \\
    \includegraphics[width=.45\columnwidth]{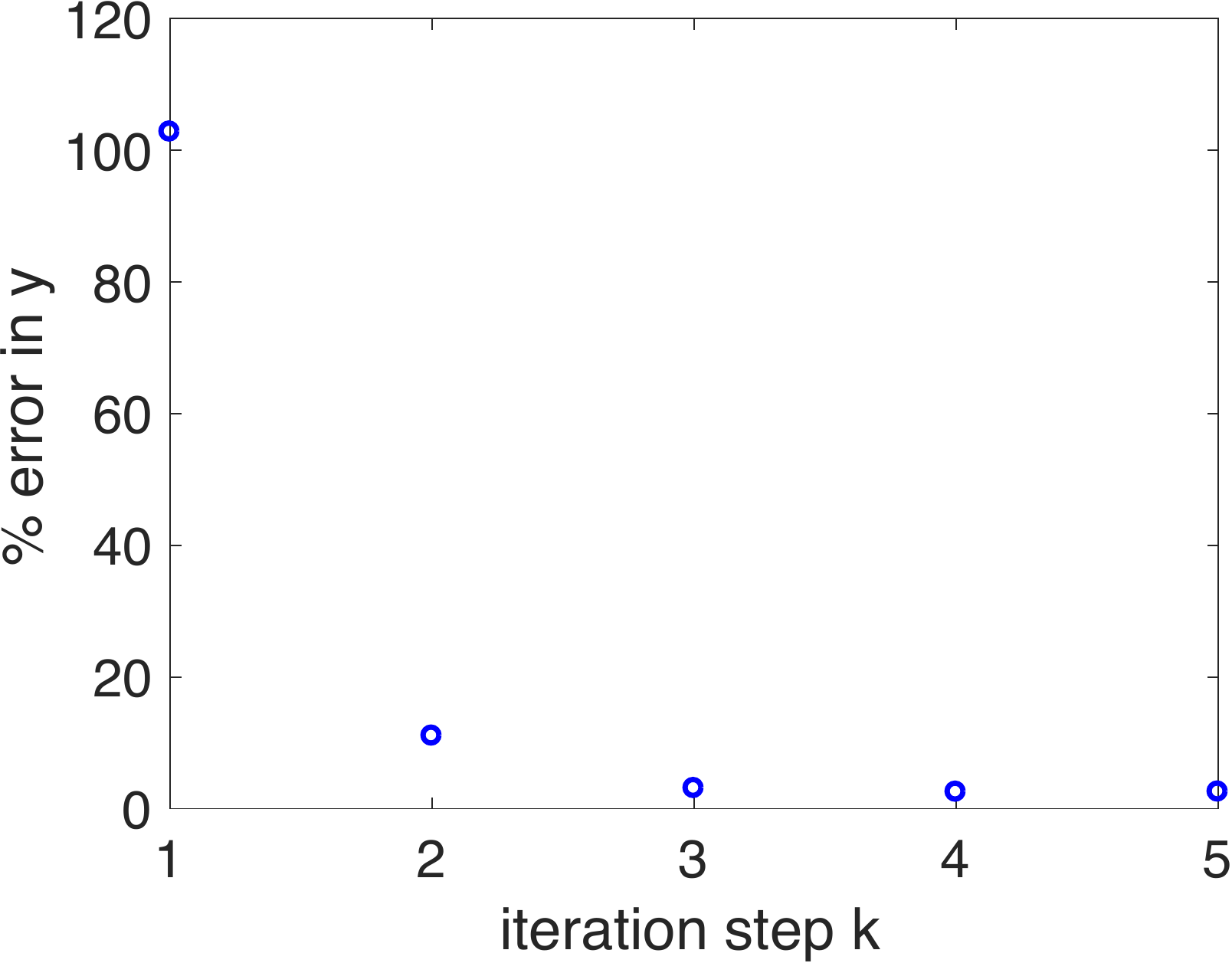} & 
    \includegraphics[width=.45\columnwidth]{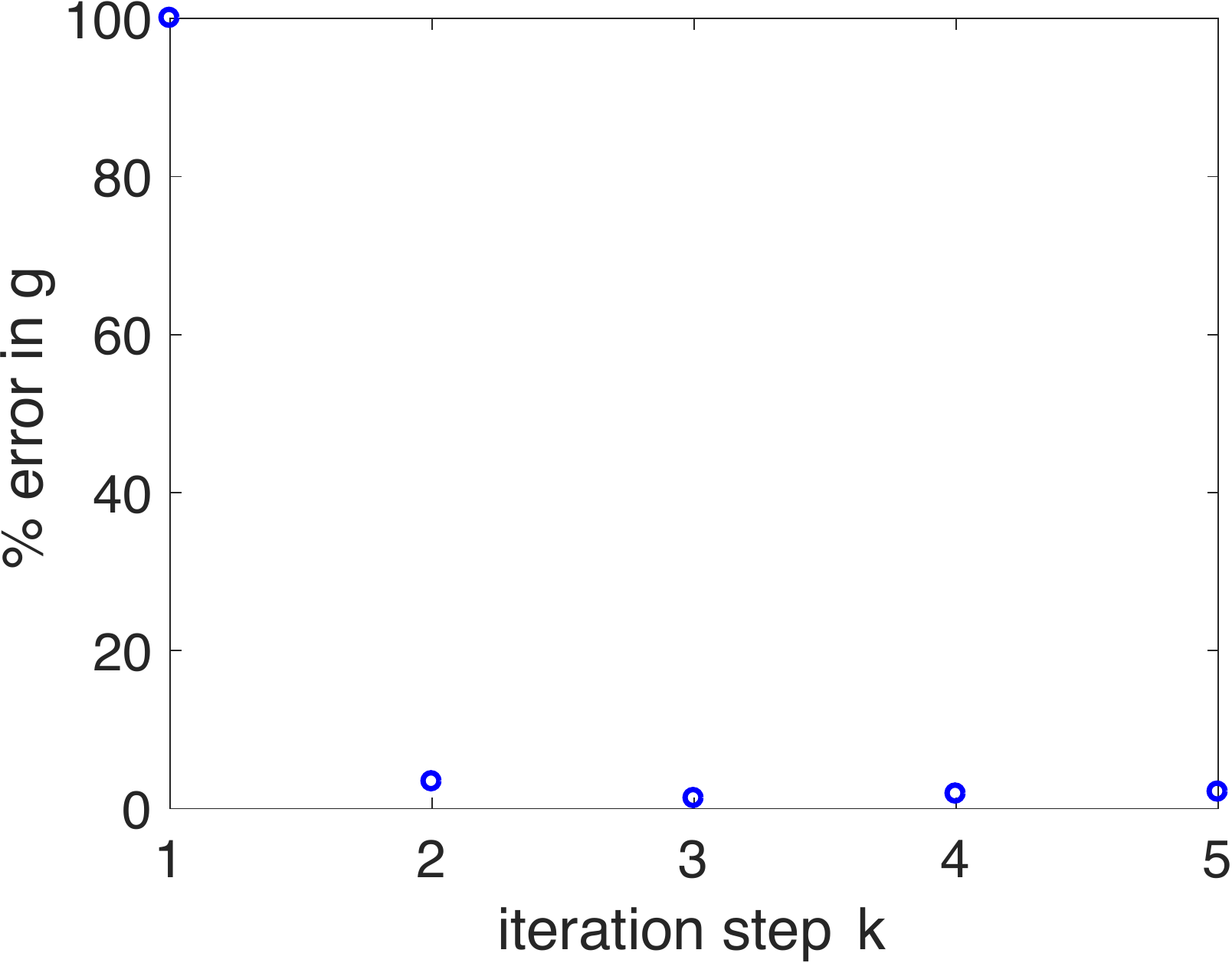} 
  \end{tabular}
  \caption{{\em{ 
(Top left) Convergence of the measured output $y_{m,k}$ to the desired output $y_d$. (Top right) Convergence of the input $u_k$ to the inverse feedforward input 
$u_d$ in Eq.~\eqref{inverse_system_eq}. (Bottom left) Output tracking error  $e_{y,k}$ from Eq.~\eqref{errors_output_model_1}. (Bottom right) 
Modeling error  $e_{g,k}$ from Eq.~\eqref{errors_output_model_2}. 
}}}
\label{Fig_5_convergence}
\end{figure}

Convergence is achieved as the iteration steps $k$ increase, i.e., the 
model  converges to the system response $\hat{g}_k\jw  \rightarrow g\jw $  and the output converges to the desired output $y_k(\cdot ) \rightarrow y_d(\cdot )$,  as seen in Fig.~\ref{Fig_5_convergence}.
The modeling error $  e_{g,k}$ given by 
\begin{equation}
\label{errors_output_model_2}
\begin{aligned}
  e_{g,k} ~ & =    \frac{ \max_{\omega\le\omega_c} | \hat{g}_k\jw-  g\jw |}{\max_{\omega\le\omega_c} | g\jw |}  \times 100,  \\
\end{aligned}
\end{equation}
reduced to $  e_{g,k^*} = 2.0$\% at the final  iteration $k=k^*$.
The  output tracking error $  e_{y,k}$ given by 
\begin{equation}
\label{errors_output_model_1}
\begin{aligned}
  e_{y,k} ~ & =    \frac{ \max_{t} |y_k(t)-  y_d(t)|}{\max_{t} |y_d(t)|}  \times 100,  \\
\end{aligned}
\end{equation}
reduced to $  e_{y,k^*} = 2.48$\% at the final  iteration $k=k^*$, which is the same as the noise level, i.e., 
\begin{equation}
  \frac{ \max_{t} |n_{y,k^*}(t)|}{\max_{t} |y_d(t)|}  \times 100  =2.48 . 
  \end{equation}

%

\vspace{0.1in}
\subsection{Impact of model learning for new output}
\begin{figure}[htb]
\centering
    \begin{tabular}{@{}cc@{}}
  \includegraphics[width=.45\columnwidth]{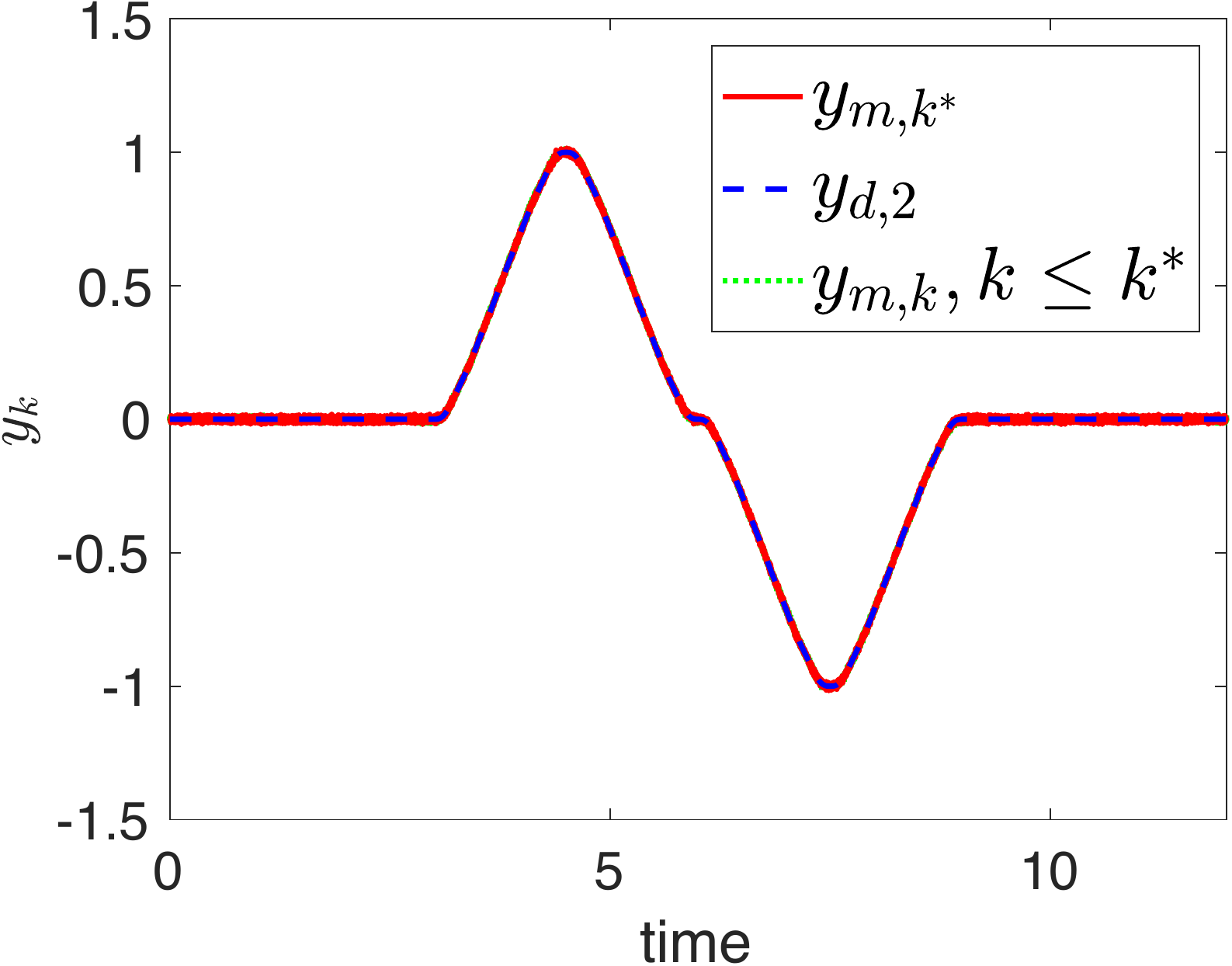} & 
    \includegraphics[width=.45\columnwidth]{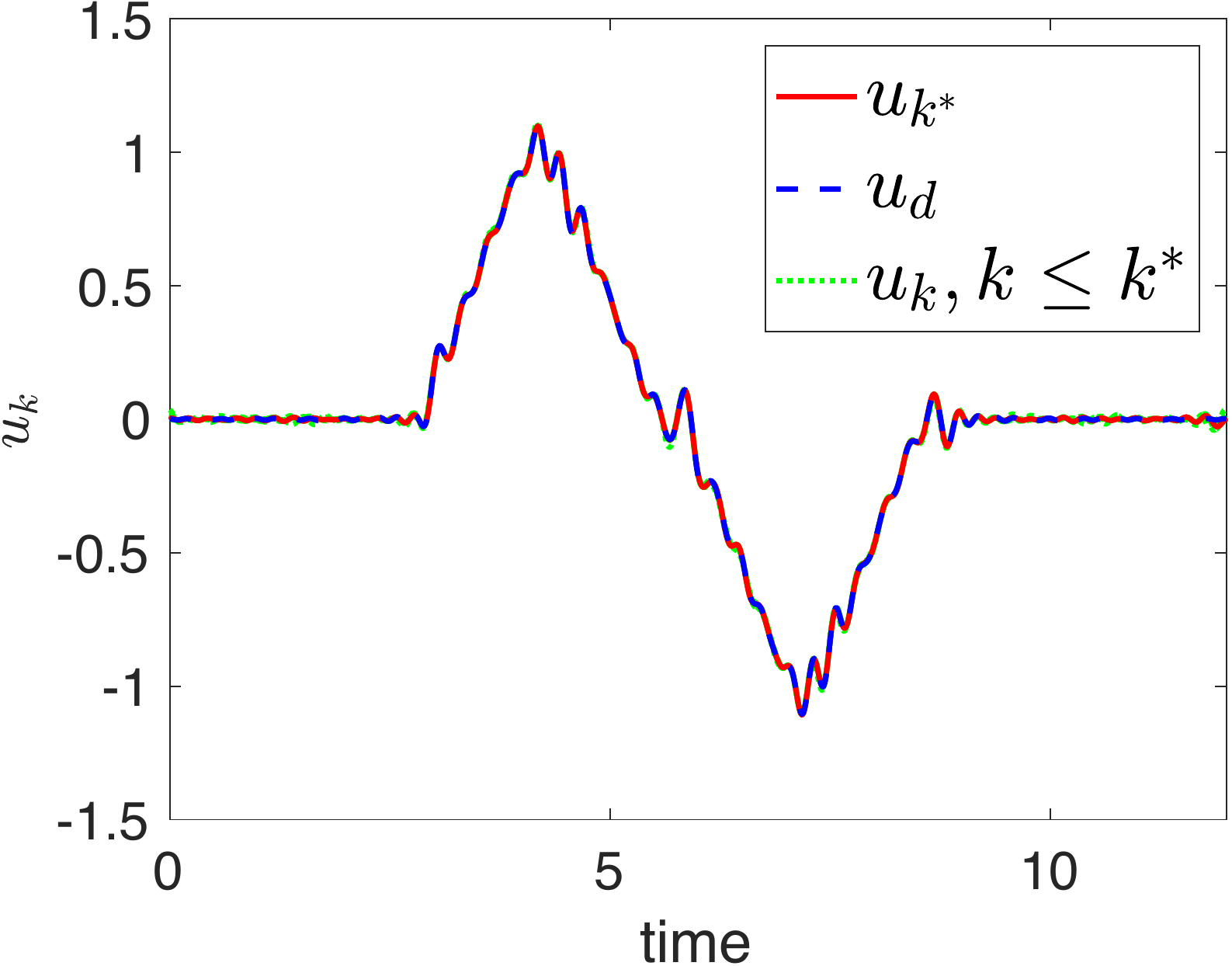}  
  \end{tabular}
  \caption{{\em{ Iteration results for new desired output $y_{d,2}$ for five iterations, $k^*=5$.
(Left) All five measured outputs $y_{m,k}$ ($k=1, \hdots 5$) overlap the desired output $y_{d,2}$. (Right) All five computed inputs $u_k$ are close to the inverse feedforward input 
$u_d$ in Eq.~\eqref{inverse_system_eq} and have substantial higher frequency content when compared to the input for the initial output trajectory $y_d$ in Fig.~\ref{Fig_5_convergence}. 
}}}
\label{Fig_6_convergence}
\end{figure}

The impact of model learning for tracking was evaluated for a new  desired output $y_{d,2}$ with a higher main 
frequency $\omega_* =  2/3$ Hz compared to the previous desired output, 
and five harmonics $N=5$ in Eq.~\eqref{desired_output_eq}.  
A plot of the desired output is shown in Fig.~\ref{Fig_6_convergence}. 
Even with this increase in the  amplitude of the desired output at higher frequencies, the 
initial input $u_1$ 
\begin{equation}
\label{iteration_law_eq_noise_3_case_2}
\begin{aligned}
u_{1}\jw ~& =   \hat{g}_1^{-1} \jw \left[   y_d\jw \right] 
~~ \forall ~ \omega \le \omega_c, 
\end{aligned}
\end{equation}
found with the initial 
model $\hat{g}_1$ 
defined as the final model  $ \hat{g}_{k^*}$ from the previous iterations,  led to good initial tracking as seen in 
Fig.~\ref{Fig_6_convergence}. (The input augmentation was not added to this initial input to clarify 
the impact of using the model from the previous iterations.) 
The input augmentation was used for iteration steps $k=2, k=3$ and the total number of iterations was 
$k^* = 5$. 
The output tracking error (with five iterations) is small --- the 
measured outputs $y_{m,k}$ for all five iterations tend to overlap the desired output $y_{d,2}$ in Fig.~\ref{Fig_6_convergence}. 
The initial tracking error  $e_{y,1}$  for the new output trajectory $y_{d,2}$ at the first iteration step was $2.97$\%, 
which  is close to the noise level of $2.12$\% ---
the tracking error $e_{y,5}$  at the end of five iterations for the second output  $y_{d,2}$ is $2.15$\%. 
Thus, as expected, the learning of the model during the iteration process with the previous desired output trajectory $y_d$ 
leads to small initial error in the iterations when learning the new  desired output $y_{d,2}$.  In this sense, the propose IML 
improves the performance and portability of the frequency-domain iterative control. 

%
%

\vspace{0.1in}
\section{Conclusions}
\label{sec_conclusions}
This article proposed a frequency-domain iterative machine learning (IML)   
approach to simultaneously update the system  model  while learning the inverse input.
Additionally,  inputs with persistency of excitation were proposed  to promote 
learning of the model by generating data at frequencies outside the main frequency content of 
the specified desired output. 
The method was applied to a simulation example, and results show the model learning  
can substantially reduce the initial error for other desired output trajectories.

\vspace{0.1in}
\section{Acknowledgment}
This work was partially supported by NSF grant CMII 1536306 and the Helen R. Whiteley Center at Friday Harbor.

\end{document}